\documentclass[10pt,preprintnumbers,nofootinbib]{revtex4-2}%
\usepackage[latin9]{inputenc}
\usepackage{amsmath}
\usepackage{amssymb}
\usepackage{dcolumn}% Align table columns on decimal point
\usepackage{graphicx}
\usepackage[colorlinks=true, pdfstartview=FitV, linkcolor=blue, citecolor=red, urlcolor=magenta]{hyperref}
\usepackage{amsfonts}
\usepackage{subfigure}
\usepackage{cleveref}
\usepackage{listings}
\usepackage{xcolor}
\usepackage{array}
\usepackage{url}
\usepackage{color}%
\usepackage{tikz}
\usepackage{pgfplots}
\usepackage{todonotes}
\usepackage{multirow}
\usepackage[outdir=./Figures]{epstopdf}

\newcommand{\oo}{\mathring}

\usepackage{booktabs} 
\newcommand{\qbar}{\raisebox{-1.5ex}[\height][0pt]{$\mathchar'26$}\mkern-9mu q}

\DeclareMathOperator{\cotan}{cotan}

\usepackage{xcolor}

\graphicspath{{Figures/}}

\begin{document}

\begin{flushleft}
\texttt{RBI-ThPhys-2024-16}\\
\texttt{ZTF-EP-24-07}
\end{flushleft}

\title{Noncommutative quasinormal modes of Schwarzschild black hole\\}

\author{Nikola Herceg}
\email{nherceg@irb.hr}
\author{Tajron Juri\'c}
\email{tjuric@irb.hr}
\author{A. Naveena Kumara }
\email{nathith@irb.hr}
%\naveencroatia
\author{Andjelo Samsarov}
\email{asamsarov@irb.hr}
\affiliation{Rudjer Bo\v{s}kovi\'c Institute, Bijeni\v cka  c.54, HR-10002 Zagreb, Croatia}

\author{Ivica Smoli\'c}
\email{ismolic@phy.hr}
\affiliation{Department of Physics, Faculty of Science, University of Zagreb, Bijeni\v cka cesta 32, 10000 Zagreb, Croatia}
\date{\today}

\begin{abstract}

We study gravitational perturbations of the Schwarzschild metric in the context of noncommutative gravity. $r-\varphi$ and $r-t$ noncommutativity are introduced through a Moyal twist of the Hopf algebra of diffeomorphisms. Differential geometric structures such as curvature tensors are also twisted. Noncommutative equations of motion are derived from the recently proposed NC vacuum Einstein equation.
Here, in addition to previously calculated axial NC potential, we present the polar solution which generalizes the work done by Zerilli.
Quasinormal mode frequencies of the two potentials are calculated using three methods: WKB, P\"oschl-Teller and Rosen-Morse.
Notably, we apply the WKB method up to the 13th order and determine the optimal order for each noncommutative parameter value individually.
Additionally, we provide comprehensive error estimations for the higher-order WKB calculations, offering insights into the accuracy of our results.
By comparing the spectra, we conclude that the classical isospectrality of axial and polar modes is broken upon spacetime quantization. Isospectrality is restored in the eikonal limit.

\end{abstract}

\maketitle

\section{Introduction}

Since its inception, general theory of relativity (GR) has proven  to be  the most successful theory of gravity.  In a wide range of different physical circumstances in which it was tested, it has prevailed against all challenges. Indeed, the theory has  passed many experimental tests involving  diverse astrophysical phenomena ranging from gravitational lensing to compact binary coalescences. However, the ranges in which gravity was tested so far     included mostly  the   weak-field,  low-speed and linear gravity  limits, while the extreme regimes like strong gravity and high curvature remained largely uninvestigated. The latter could presumably be accessed by studying the near-horizon region of black holes where energies as large as the Planck scale are expected to be present.  \\

At this energy scale, the gravitational effects become comparable to  quantum effects and  one might expect a modification of GR to be triggered there. 
 This is one of those points where the gravitational waves   come into play.
Besides them being one of the most important predictions of general relativity, gravitational waves
 appear to play a crucial role in accessing  the behaviour of GR under extreme conditions, i.e. in a regime where GR  is not   well enough studied yet.
Namely, due to the fact that gravitational waves originate from regions of spacetime where gravity is extremely strong, just like the near-horizon region of black holes, they have emerged as highly suitable candidates for testing GR under such extreme conditions.
In fact, since their recent experimental discovery, gravitational waves have been established as one of the most  powerful tools  for  studying extreme  regimes  of relativistic speeds, large curvature,  and strong gravity. In this way, gravitational waves provided a promising way for exploring spacetime dynamics in  regions   previously  inaccessible to an experimental  enquiry  and  helped  bring  them   under a direct experimentalist's gaze. \\

Gravitational waves as observed by the current ground-based gravitational wave detectors have been extensively used for testing various aspects of general relativity, but  no discrepancies from the theory have been reported so far \cite{LIGOScientific:2016lio,LIGOScientific:2017vwq,LIGOScientific:2016aoc}. As it currently appears, possible deviations from GR, if anywhere found in the literature, may most likely  be attributed to  a false identification of  GR violation arising from a series of technical issues such as detector noise, signal overlaps, gaps in the data, detector calibration,  source misidentification or others \cite{Gupta:2024gun}. This motivates an ongoing and increasingly enthusiastic search for potential deviations in the field. \\

However, strong gravity regimes, such as those found in the vicinity of a black hole, may present a more challenging scenario for general relativity, as its validity could be subjected to a more rigorous examination in these contexts.
As already stated, the region near a black hole horizon admits access to energies of the order of the Planck scale. As a result, GR may be potentially modified there, causing that some of its predictions, including gravitational QNM spectrum of black holes, may be altered too \cite{Barausse:2014tra,Cardoso:2016rao,Zhao:2023uam,Campos:2021sff,Heidari:2023egu,Konoplya:2019xmn}. 
Indeed, the latter scenario receives  significant support within a scope of various quantum gravity  theories, which suggest that strong gravity and high energy regimes like those found in a near-horizon region of black hole  may create environment where the usual point-like structure of spacetime becomes replaced with certain noncommutative geometry constructs \cite{Doplicher:1994zv, Doplicher:1994tu} ultimately  affecting the relaxation dynamics of  black holes and their related ringdown spectra. 
Even more striking evidence of the discrete nature of spacetime and a possible necessity for modifying classical GR might be found in phenomena like gravitational wave echoes or tidal heating \cite{Agullo:2020hxe}. 
  Alternative theories of gravity -- such as the Einstein-Aether model \cite{Konoplya:2006rv}, Einstein-power-Maxwell scenarios \cite{Rincon:2018sgd, Panotopoulos:2017hns} and studies of acoustic black holes \cite{Vieira:2021ozg} -- have similarly revealed that deviations from the classical GR QNM spectra may occur.\\

The initial steps toward  quantifying the black hole ringdown spectrum have been carried out on the example of the Schwarzschild black hole by Regge and Wheeler for the odd-parity (axial) perturbations \cite{Regge:1957td} and subsequently by Zerilli for  the even-parity (polar) perturbations \cite{Zerilli:1970se}. In both cases the problem was shown to reduce to a scattering problem at a specially devised  potential barrier, thus establishing the full analogy between  the theory of scattering of gravitational waves on a background of a black hole and a scattering theory in quantum mechanics \cite{Vishveshwara:1970zz,Thorne:1980ru}. The outcome of the analysis was a master equation in a form of a 1-dim Schr\"odinger-like equation in a radial coordinate with a particular effective potential that came out  after  separating  the angular variables in the wave equation, using tensor spherical harmonics with angular indices $(\ell,m)$. Careful study of the master equation  revealed that quasinormal modes in classical general  relativity have the property of isospectrality.  In particular, the axial and polar perturbations of the Schwarzschild black hole were shown by Chandrasekhar and Detweiler to have identical spectra \cite{Chandrasekhar:1975zza}. \\

In our previous papers \cite{Herceg:2023zlk, Herceg:2023pmc, Herceg:2023lmt}.  we have set up a noncommutative differential geometry framework sufficiently suited for studying metric perturbations in noncommutative gravity. This formalism is based on considering  a set of deformed  diffeomorphisms encompassing  the symmetries of a noncommutative space which conveniently  find their description  within the structure of Hopf algebra. In addition, the formalism was applied to obtain noncommutative corrections to the Regge-Wheeler potential that governs the axial perturbations of the Schwarzschild black hole. \\

In the present paper, we continue on our previous work by presenting noncommutative gravitational perturbations of the Schwarzschild black hole in their entirety, giving a description of metric perturbations in noncommutative gravity for both axial and polar cases. We do this by first recapitulating the results for axial perturbations  obtained previously and then by extending the analysis to include polar perturbations. Curiously, contrary to the classical case, we find that these two types of perturbations do not share the same QNM spectra, yielding a conclusion that the presence of quantum spacetime violates isospectrality. \\

The plan of the paper is the following. In Section \ref{secII} we briefly present a formalism of NC differential geometry  and explain how it can be applied to  linearized gravitational perturbation theory. In Section \ref{secIII} we single out the NC corrections to the metric perturbations around the Schwarzschild black hole and calculate  noncommutative corrections to the equation of motion for axial (i.e. odd-parity) and polar (i.e. even-parity) metric perturbations. In Section \ref{sec3A} we discuss the axial sector. Rephrasing the problem in terms of a Schr\"odinger-type equation enables us to identify the NC correction to the Regge-Wheeler potential. Section \ref{sec3B} is devoted to an extensive and in-detail analysis of the noncommutative polar perturbations of the Schwarzschild background. Utilising the classical infinitesimal diffeo-transformations and a series of field redefinitions, as well as coordinate transformations we were able to determine the NC correction to the equation of motion governing the NC  polar metric perturbations and to extract the related quantum correction to the Zerilli potential.  \\

In Section \ref{secIV} we gain the first insight into the spectra by applying the WKB, P\"oschl-Teller and Rosen-Morse methods.
In Section \ref{secV} we demonstrate that the isospectrality between two types of perturbations no longer holds, in contrast to a  well-established result in classical GR. We come to this conclusion  by arguing that the two potentials may not be connected through a Darboux transformation, making the criterion for equivalence between the two potentials impossible to fulfill. This conclusion has been  further corroborated by a detailed semi-analytic analysis of the QNM spectra in both cases, with the analysis being based on the WKB method \cite{Schutz:1985km} up to 13th order in the calculation. The paper ends with  concluding remarks and 3 appendices where we present some technical details.
In Appendix \ref{Zerilli_gauge_app} we present the usual Zerilli gauge for the polar perturbation sector.
In Appendix \ref{app-coeff} we write out the coefficients of transformations used for finding the NC corrections to the polar sector.
In Appendix \ref{appc} we give the QNM frequencies obtained through the order-optimized WKB method together with error estimations up to the 13th order.

\section{Noncommutative differential geometry} \label{secII}

We consider a noncommutative gravity theory constructed using the mathematical framework of Hopf algebra theory \cite{Aschieri:2005yw, Aschieri:2005zs, Aschieri:2009qh}. First, we will provide a clear and concise introduction to Hopf algebras and their Drinfeld twist deformations. 
In the second subsection we outline the main building blocks of NC differential geometry - a theory that arises naturally by imposing covariance under the Hopf algebra of deformed diffeomorphisms. 
For a more detailed and pedagogical introduction to the topic, see \cite{Schenkel:2011biz}.

\subsection{Hopf algebra and deformed diffeomorphisms}

The Lie algebra of vector fields $(\Xi, [\cdot , \cdot] )$ plays a vital role in differential geometry. It describes the infinitesimal diffeomorphisms of the manifold $\mathcal{M}$. It acts on tensor fields via the Lie derivative $\pounds$. From the Lie algebra $(\Xi, [\cdot, \cdot])$, we can create a universal enveloping algebra $U\Xi$, which we can upgrade to a Hopf algebra. 
Universal enveloping algebra is a free algebra on generators of the Lie algebra $\Xi$ with associative product $\mu : U\Xi \otimes U\Xi \to U\Xi$ and unit $\eta : \mathbb{C} \to U\Xi$  obtained by quotienting the former (the free algebra) by an ideal generated by relations $ab - ba - [a,b] = 0$.  \\

To encode the intuitive concepts of Leibniz rule, inverse, and normalization that we have on the Hopf algebra, we use $\mathbb{C}$-linear maps coproduct $\Delta: U\Xi \rightarrow U\Xi \otimes U\Xi$, antipode $S: U\Xi \rightarrow U\Xi$ and counit $\epsilon: U\Xi \rightarrow \mathbb{C}$. While $\Delta$ and $\epsilon$ are multiplicative maps, the map $S$ associated with inversion in the corresponding Lie group is antimultiplicative. The object of interest here is the structure $H = (U\Xi, \mu, \eta, \Delta, \epsilon, S)$, known as a Hopf algebra.  \\

A Hopf algebra can be understood as a generalization of the enveloping algebra that includes three important maps: coproduct $\Delta$, counit $\epsilon$, and antipode $S$. These maps must satisfy certain compatibility criteria. To be considered a Hopf algebra, it must meet the following three conditions for all $\xi \in H$:
\begin{equation}
    \begin{aligned}
        (\Delta \otimes \text{id})\Delta(\xi)=&\ (\text{id} \otimes \Delta)\Delta (\xi),\\
        (\epsilon \otimes \text{id})\Delta(\xi)=&\  \xi=(\text{id} \otimes \epsilon)\Delta (\xi),\\
       \mu ( (S \otimes \text{id})\Delta(\xi)=&\ \epsilon (\xi)1=\mu((\text{id} \otimes S)\Delta (\xi)).
    \end{aligned}
\end{equation}
In summary, the Lie algebra of diffeomorphisms $(\Xi, [\cdot , \cdot] )$ can be embedded into a Hopf algebra $H = (U\Xi, \mu, \eta, \Delta, \epsilon, S)$. This Hopf algebra is known as the Hopf algebra of diffeomorphisms. It contains the relevant information on the Leibniz rule (through the coproduct $\Delta$), the inverse of a diffeomorphism (through the antipode $S$), and the normalization process (through the counit $\epsilon$). \\

Let us present some properties of the Hopf algebra $H$. One important concept is that of a Drinfeld twist. A Drinfeld twist refers to an invertible element $\mathcal{F}\in H \otimes H$ that satisfies the following two conditions:
\begin{equation}
    \begin{aligned}
	    (\mathcal{F}\otimes 1) (\Delta \otimes \text{id})\mathcal{F}=& \ (1\otimes \mathcal{F}) (\text{id}\otimes \Delta)\mathcal{F},\\
        (\epsilon \otimes \text{id})\mathcal{F}=& \ 1=(\text{id}\otimes \epsilon)\mathcal{F}.
    \end{aligned}
\end{equation}
Notation $\mathcal{F} = f^\alpha \otimes f_\alpha, \: \mathcal{F}^{-1} = \bar{f}^\alpha \otimes \bar{f}_\alpha$, where summation over $\alpha$ is assumed, will often be employed.
We further require that $\mathcal{F}=1\otimes 1+\mathcal{O}(a)$, where $a$ represents a deformation parameter, in order to keep the zeroth order (classical limit) unchanged.\footnote{To implement this in practice, we need to extend the underlying field $\mathbb{C}$ to a formal power series in deformation parameter $a$. For details refer to \cite{Schenkel:2011biz}.} If a Drinfeld twist $\mathcal{F}$ of the Hopf algebra $H$ is provided, it is well-known that we can build a new Hopf algebra $H^{\mathcal{F}}:= (U \Xi, \mu, \eta, \Delta^{\mathcal{F}}, \epsilon, S^{\mathcal{F}})$ by twisting the coproduct and the antipode as follows:

\begin{equation}
    \begin{aligned}
        \Delta ^{\mathcal{F}}(\xi):=&\ \mathcal{F} \Delta(\xi) \mathcal{F}^{-1},\\
         S ^{\mathcal{F}}(\xi):=&\ \chi S(\xi) \chi ^{-1}.
    \end{aligned}
\end{equation}
Here, $\chi := f ^\alpha S(f_\alpha)$ and $\chi ^{-1}:=  S(\bar f^\alpha) \bar f _\alpha$. Note that $f ^\alpha, f_\alpha$, $\bar f_\alpha$, $\bar f ^\alpha$ are all elements in $H$. 
\\ \par
The twisted Hopf algebra of diffeomorphisms, denoted by $H^{\mathcal{F}}$, raises the question of whether and in what sense it differs from the original Hopf algebra $H$. 
It should be noted that $H$ is a cocommutative\footnote{The Hopf algebra of diffeomorphisms has a unique property called cocommutativity. It means that the coopposite coproduct of any generic element $\xi$ in $H$, denoted as $\Delta ^{\text{cop}} (\xi) = \xi _2 \otimes \xi _1$, is equal to the coproduct $\Delta (\xi) = \xi_1 \otimes \xi_2 = \xi \otimes 1 +1\otimes \xi$ itself, where we used the Sweedler notation for the coproduct. As a result, the Hopf algebra $H$ is referred to as a cocommutative Hopf algebra.} Hopf algebra, i.e. $\Delta ^{\text{cop}}=\Delta$. It is easy to show that the coopposite twisted coproduct $(\Delta^\mathcal{F})^{\text{cop}}$ is related to the twisted coproduct $\Delta^\mathcal{F}$ by performing a conjugation with the help of the element $\mathcal{R} =\mathcal{F}_{21} \mathcal{F}^{-1}\in  H^{\mathcal{F}} \otimes H^{\mathcal{F}}$ for all $\xi \in $ $H^{\mathcal{F}}$ (where $\mathcal{F}_{21}=f_\alpha \otimes f^{\alpha}$ and  $\mathcal{F}_{21}^{-1}=\bar f_\alpha \otimes \bar f^{\alpha}$), as given by the equation:
\begin{equation} \label{cop}
    (\Delta ^\mathcal{F})^{\text{cop}}(\xi) = \mathcal{R} \Delta ^{\mathcal{F}} (\xi) \mathcal{R}^{-1}.
\end{equation}
This element $\mathcal{R}$ is called a universal $\mathcal{R}$-matrix and satisfies the quantum Yang-Baxter equation, giving rise to braiding. We use the notation $\mathcal{R} = R^\alpha \otimes R _\alpha$ and $\mathcal{R}^{-1} =\bar R ^\alpha \otimes \bar R _\alpha $ (sum over $\alpha$ assumed) for the $\mathcal{R}$-matrix and its inverse, respectively. \\

The twisted Hopf algebra of diffeomorphisms  $H^{\mathcal{F}}$ is generally not cocommutative. This means that $H^{\mathcal{F}}$ differs structurally from the Hopf algebras generated by Lie algebras via the universal enveloping algebra construction. As a result of this noncocommutative behavior of $H^{\mathcal{F}}$, we will obtain a noncommutative structure on the spaces that the Hopf algebra acts on (modules), such as the algebra of functions on the spacetime manifold $\mathcal{M}$. \\

In the context of noncommutative manifolds, it is important to consider the construction of scalar, vector, and tensor transformations that are compatible with the twisted Hopf algebra of diffeomorphisms. Our focus is on the Hopf algebra $H^{\mathcal{F}}$ which characterizes the deformed infinitesimal diffeomorphisms that correspond to the symmetry of a noncommutative manifold. To illustrate, let us consider the simplest type of tensor field: the smooth and complex functions $\mathcal{C}^\infty (\mathcal{M})$. This space can be equipped with an algebra structure by employing the pointwise multiplication $(h k)(x) = h(x) k(x)$, for all $h, k \in \mathcal{C}^\infty (\mathcal{M})$. The algebra $\mathcal{A} := (\mathcal{C}^\infty (\mathcal{M}), \cdot)$ is covariant under the Hopf algebra $H$, meaning $\xi \rhd \cdot  \ (f \otimes g) = \cdot \ \Delta \xi \rhd (f \otimes g)$ for $\xi \in H, \ f, g \in C^\infty(M)$. This is the usual Leibniz rule for the action of the Lie derivative, or more formally, left $H$-module algebra property of $\mathcal A$.  \\

However, when it comes to nontrivial deformations generated by $\mathcal{F}$, $\Delta$ in the preceding relation changes to $\Delta^\mathcal{F}$ and algebra $\mathcal{A}$ fails to be covariant under $H^{\mathcal{F}}$. 
To resolve this issue, we must deform the product in $\mathcal A$ according to $\mathcal{F}$, transforming $\mathcal{A}$ into $\mathcal{A}_\star$ -- the algebra covariant under $H^{\mathcal{F}}$. In other words, the twisted Hopf algebra $H^{\mathcal{F}}$ describes the symmetries of a noncommutative manifold underlying the algebra $\mathcal{A}_\star = (\mathcal{C}^\infty (\mathcal{M}), \star)$, where the deformed multiplication, the so-called  $\star$-product, is given by:
\begin{equation} \label{genericstar}
    h \star k := \cdot \  \mathcal{F}^{-1} (h \otimes k)= \bar f^\alpha (h) \bar f_\alpha (k),
\end{equation}
for all $h, k \in \mathcal{C}^\infty (\mathcal{M})$. A similar prescription can be used to deform the tensor product from $\otimes_\mathcal{A}$ to $\otimes_\mathcal{A_\star}$ and space of tensors from $\mathcal{T}$ to $\mathcal{T}_\star$. It is important to note that as a vector space $\mathcal{A}$ is isomorphic to $\mathcal{A}_\star$ and also $\mathcal{T}$ is isomorphic to $\mathcal{T}_\star$, therefore it is only the algebraic structure which we deform.
Property \eqref{cop} of the coproduct is reflected inside the algebra ${\mathcal A}_\star$ as braided commutativity of $\mathcal{A}_\star$:
\begin{equation}
	f \star g = \bar R^\alpha(g) \star \bar R_\alpha(f).
\end{equation}

To implement noncommutative gravity practically, we require a specific Hopf algebra of deformed diffeomorphisms. 
More precisely, for our study, we select a particular twist $\mathcal F$, known  as the Moyal-Weyl twist, and utilize it to twist $H$.
If we consider $\mathcal{M}=\mathbb{R}^N$ and use $x^\mu$ ($\mu=1,...,N$) as local coordinate functions on $\mathcal{M}$, then the derivatives $\partial _\mu$ along $x^\mu$ provide a local basis of $\Xi$. This means that any vector field $v\in \Xi$ can be expressed as $v=v^\mu (x) \partial _\mu$, where the coefficient functions $v^\mu \in C^\infty(\mathcal{M})$. Notably, $\partial _\mu \in \Xi$ are locally defined vector fields for all $\mu = 1, . . . , N$. The Moyal-Weyl twist $\mathcal{F}$ is an element in $U\Xi \otimes U\Xi$, and is defined by the following expression:
\begin{equation} \label{Moyal-Weyl}
    \mathcal{F}=\exp \left( -i\Theta ^{\mu \nu} \partial _\mu \otimes \partial _\nu  \right),
\end{equation}
where $\Theta ^{\mu \nu}$ is a constant and antisymmetric matrix.\\

\subsection{Quantum Lie algebras and NC geometry}

We can associate a quantum Lie algebra ($\Xi, [\cdot, \cdot]_\star$) to $H^{\mathcal{F}}$, similar to how we can associate the Lie algebra $(\Xi, [\cdot, \cdot])$ of vector fields to $H$. However, this quantum Lie algebra fits naturally inside another Hopf algebra $H_\star$, which is isomorphic to $H^{\mathcal F}$. Now we describe its construction in brief, with universal enveloping algebras as the starting point. \\

First note that $(U\Xi, \mu)$ is a left $H$-module algebra via the adjoint action. We deform this module algebra to $(U\Xi, \mu_\star)$ by twisting the associative product $\mu$. There is a remarkable relation between the algebras $(U\Xi, \mu)$ and $(U\Xi, \mu_\star)$, as they are isomorphic. The algebra isomorphism $D : (U\Xi, \mu) \rightarrow (U\Xi, \mu_\star)$ is given by the adjoint action $D(\xi):=
\text{Ad}_{\bar f^\alpha }(\xi) \bar f_\alpha$ for all $\xi \in U\Xi$, where $\text{Ad}_{\bar f^\alpha} (\xi) := \mu (1 \otimes S) \Delta {\overline f}^\alpha = \bar f^\alpha_1  \xi  S(\bar f^\alpha_2)$.  This algebra isomorphism can be lifted to the Hopf algebra level by introducing $H_\star = (U\Xi, \mu_\star, \eta, \Delta_\star, \epsilon, S_\star)$, which is isomorphic to $H^{\mathcal{F}}$. The quantum Lie algebra and the $R$-matrix it carries are of central importance in our approach; details of the co-algebraic  structures of $H_\star$ can be found in \cite{Aschieri:2009qh,Schenkel:2011biz}. \\

Since $H_\star$ and $H^{\mathcal{F}}$ are isomorphic Hopf algebras, any representation of $H^{\mathcal{F}}$ is also a representation of $H_\star$. We perform the $H_\star$-action by the $\star$-Lie derivative, for all $\xi \in U\Xi$, as follows:
\begin{equation}
    \pounds_\xi ^\star := \pounds_{D(\xi)}.
\end{equation}
When acting with the $\star$-Lie derivative on a $\star$-tensor product of tensor fields, we obtain
\begin{equation}
    \pounds_v ^\star (\tau \otimes _{A_\star} \sigma )=\pounds_v ^\star (\tau) \otimes _{A_\star} \sigma + \bar R ^\alpha (\tau) \otimes _{A_\star} \pounds_{\bar R_\alpha (v)}^\star (\sigma),
\end{equation}
for all $v\in \Xi$, and for all tensors $\tau, \sigma \in \mathcal{T_\star}$. 
Furthermore, we can construct a quantum Lie algebra $(\Xi,[\cdot, \cdot]_\star)$, where the $\star $-Lie bracket $[\cdot, \cdot]_\star$ satisfies the deformed antisymmetry property:
\begin{equation}
    [v,w]_\star = -[\bar R^\alpha (w), \bar R_\alpha (v)]_\star,
\end{equation}
and the deformed Jacobi identity:
\begin{equation}
	[v,[w,z]_\star]_\star =  [[v,w]_\star,z]_\star  + [\bar R^\alpha (w),[ \bar R_\alpha (v),z]_\star]_\star,
\end{equation}
for all $v, w, z \in \Xi$. The quantum Lie algebra $(\Xi,[\cdot, \cdot]_\star)$ represents the infinitesimal deformed diffeomorphisms. \\

%\subsection{covariant derivative and curvature}

In the framework of noncommutative differential geometry, we can introduce the covariant derivative, torsion, and curvature in a natural way.  A $\star$-covariant derivative $\hat{\nabla} _v$ is introduced along a vector field $v\in \Xi$. It is a $\mathbb{C}$-linear map $\hat{\nabla}_v : \Xi \rightarrow \Xi$ that satisfies the following conditions: 
\begin{equation}
\begin{aligned}
       \hat{\nabla}  _{v+w} \: z=& \: \hat{\nabla}  _v z+\hat{\nabla}  _w z ,\\
	\hat{\nabla}  _{h\star v} \: z=& \: h \star \hat{\nabla}  _v z, \\
        \hat{\nabla}  _{v}(h \star z)=& \: \pounds_v^\star (h) \star z+\bar R ^\alpha (h)\star \hat{\nabla}  _{\bar R_\alpha (v)}z,
\end{aligned}  
\end{equation}
for all $v, w, z \in \Xi$ and $h \in \mathcal{C}^\infty_\star (\mathcal{M})$. When a $\star$-covariant derivative is given, its $\star$-torsion and $\star$-curvature are defined by $\mathbb{C}$-linear maps $\hat{T} : \Xi \otimes \Xi \rightarrow \Xi$ and $\hat{R} : \Xi \otimes \Xi \otimes \Xi \rightarrow \Xi$, respectively.
\begin{equation} \label{NCRiemann}
    \begin{aligned}
        \hat{T}  (v,w):=& \: \hat{\nabla}  _v w-\hat{\nabla}  _{\bar R^\alpha (w)} \bar R _\alpha (v)-[v,w]_\star ,\\
        \hat{R} (v,w,z):=& \: \hat{\nabla}  _v \hat{\nabla}  _w z - \hat{\nabla}  _{\bar R^\alpha (w)} \hat{\nabla}  _{\bar R_\alpha (v)} z- \hat{\nabla}  _{[v,w]_\star}z,
    \end{aligned}
\end{equation}
for all $v,w,z \in \Xi$. To construct the $\star$-\emph{Ricci tensor} $\hat{R}  : \Xi \otimes \Xi \rightarrow \mathcal{C}^\infty_\star(\mathcal{M})$, we need a local basis and a contraction covariant under $H^{\mathcal{F}}$.\footnote{Since vector fields and one-forms are dual to each other in differential geometry, we can define a contraction $\langle \cdot, \cdot \rangle: \Xi \times \Omega ^1 \rightarrow \mathcal{C}^\infty (\mathcal{M})$. Here, $\Omega ^1$ represents the space of smooth and complex 1-forms. These contractions are $H$-covariant, which means that for all $v\in \Xi$, $\omega \in \Omega ^1$, and $\xi \in H$, we have $\pounds_\xi (\langle v, \omega \rangle )=  \langle \pounds_{\xi _1} (v) , \pounds_{\xi _2} (\omega) \rangle $. In order to make $\langle \cdot , \cdot \rangle$ covariant under $H^{\mathcal{F}}$, we use the inverse twist and define $ \langle v, \omega \rangle_\star := \langle \bar f^\alpha (v) , \bar f _\alpha (\omega) \rangle,$ for all $v\in \Xi,$ and $ \omega \in \Omega^1$.} In the $\star$-dual basis $\{dx^\nu\}$ for which $\langle \partial _\mu , dx^\nu \rangle _\star =\delta_\mu^{~\nu}$ holds, the contraction defines the $\star$-Ricci tensor as follows: 
\begin{equation} \label{NCRicci}
\hat{R}  (v,w)= \langle dx^\mu , \hat{R}  (\partial _\mu, v, w)\rangle _\star, 
\end{equation} 
for all $v, w \in \Xi$. It is independent of the choice of basis. The noncommutative Ricci tensor \eqref{NCRicci} is not symmetric, due to the fact that the noncommutative Riemann tensor defined in \eqref{NCRiemann} is $\mathcal{R}$-antisymmetric in the first two indices,
\begin{equation}
    \hat R (u, v, z)= - \hat R(\bar R^c(v) , \bar R_c(u),z ),
\end{equation}
but not in the last two. For the deformations induced by the Moyal-Weyl twist \eqref{Moyal-Weyl}, the Christoffel symbols completely specify the $\star$-covariant derivative as shown below:
\begin{equation}
    \hat{\nabla}  _{\partial _\mu} \partial _\nu =\hat \Gamma _{~\mu \nu}^{ \rho} \star \partial _\rho = \hat \Gamma _{~\mu \nu}^{\rho}  \partial _\rho.
\end{equation}
The last equality follows from the fact that the twist acts trivially on all $\partial _\rho$. In the basis $\{\partial _\rho\}$, the $\star$-torsion, $\star$-curvature and $\star$-Ricci tensor can be expressed as:
\begin{equation}
    \begin{aligned}
	    \hat{T}  (\partial _\mu , \partial _\nu)= & ~\hat T_{~\mu \nu}^\rho \partial_\rho = \left( \hat \Gamma _{~\mu \nu}^{\rho} -\hat \Gamma _{~\nu \mu}^{ \rho} \right) \partial _\rho, \\
        \hat R (\partial _\mu , \partial _\nu , \partial _\rho)= & \ \hat R_{~\mu \nu \rho}^\sigma \partial _\sigma
	    =\left(\partial _\mu \hat \Gamma _{~\nu \rho}^{ \sigma} -\partial _\nu \hat \Gamma _{~\mu \rho}^{ \sigma} + \hat \Gamma _{~\nu \rho}^{ \tau} \star \hat  \Gamma _{~\mu \tau}^{ \sigma}-\hat \Gamma _{~\mu \rho}^{ \tau} \star \hat \Gamma _{~\nu \tau}^{ \sigma} \right) \partial _\sigma ,\\
	    \hat R (\partial _\nu , \partial _\rho)=& \ \hat R_{\nu \rho}
	    =\partial _\mu \hat \Gamma _{~\nu \rho}^{ \mu} -\partial _\nu \hat \Gamma _{~\mu \rho}^{ \mu} + \hat \Gamma _{~\nu \rho}^{\tau} \star \hat \Gamma _{~\mu \tau}^{ \mu}- \hat \Gamma _{~\mu \rho}^{ \tau} \star \hat \Gamma _{~\nu \tau}^{\mu}.
    \end{aligned}
\end{equation}
It's important to note that these equations hold in the case of Abelian twist in the so-called nice basis \cite{Schenkel:2011biz}. Abelian twist means that vector fields generating the twist commute and nice basis means that vector fields generating the twist commute with the basis vectors. For the twist \eqref{Moyal-Weyl} this trivially holds since it is composed of derivatives with respect to a coordinate chart. \\

To formulate a noncommutative gravity theory, we also need to introduce a metric field $g = g^\alpha \otimes _{\mathcal{A}_\star} g_\alpha \in \Omega ^1 \otimes _{\mathcal{A}_\star} \Omega ^1 $. 
%Our requirement for this metric field is that it should be symmetric, real, and nondegenerate. 
We can define the $\star$-inverse metric $g^{-1}=g^{-1\alpha} \otimes _{\mathcal{A}_\star} g_\alpha ^{-1}
 \in \Xi  \otimes _{\mathcal{A}_\star} \Xi $ by imposing the following conditions for all $v\in \Xi$ and $\omega \in \Omega ^1$:
  \begin{equation}
     \begin{aligned}
         \langle \langle v, g\rangle _\star , g^{-1} \rangle _\star &= \langle v, g^\alpha \rangle _\star \star \langle g_\alpha , g^{-1 \beta}\rangle _\star \star g^{-1}_\beta =v,\\
         \langle \langle \omega, g^{-1}\rangle _\star , g\rangle _\star &= \langle \omega, g^{-1 \beta} \rangle _\star \star \langle g_\beta ^{-1} , g^{\alpha}\rangle _\star \star g_\alpha =\omega .
     \end{aligned}
 \end{equation}

In the context of the Moyal-Weyl twist \eqref{Moyal-Weyl}, the metric field can be expressed as $g=dx^\mu \otimes _\mathcal{A} dx^\nu g_{\mu \nu}= dx^\mu \otimes _{\mathcal{A}_\star} dx^\nu \star g_{\mu \nu}$, since the $\star$ symbol reduces to its zeroth order commutative multiplication when acting on $\partial _\rho$ or $dx^\mu$. The inverse metric field is given by $g^{-1}=g^{\star \mu \nu} \star \partial _\mu \otimes _{\mathcal{A}_\star} \partial _\nu$, where $g_{\mu \nu}$ and $g^{\star \mu \nu}$ satisfy the conditions: 
\begin{equation}
	g_{\mu \nu} \star g^{\star \nu \rho}=\delta^{~\rho} _\mu, \qquad g^{\star \mu \nu} \star g_{\nu \rho}=\delta_{~\rho} ^\mu .
\end{equation}
Specifically, we have \cite{Juric:2022bnm}
\begin{equation}
g^{\star\alpha\beta}=g^{\alpha\beta}-g^{\gamma\beta}i \Theta^{ \mu \nu }(\partial_ \mu g^{\alpha\sigma})(\partial_ \nu g_{\sigma\gamma})+\mathcal{O}(\Theta^2).
\end{equation}
The unique torsion-free, $\hat{T}=0$, and metric compatible\footnote{Covariant derivative of the metric vanishes, but covariant derivative of the metric inverse does not \cite{Aschieri:2009qh}.}, $\hat{\nabla}(g)=0$, NC Levi-Civita connection is given by \cite{Aschieri:2009qh},
\begin{equation}\label{sLC}
\hat \Gamma_{\mu\nu}^{ \rho} =\frac{1}{2}g^{\star\rho\sigma}\star\left(\partial_{\mu}g_{\nu\sigma}+\partial_{\nu}g_{\mu\sigma}-\partial_{\sigma}g_{\mu\nu}\right).
\end{equation}
In the next section, we will use this formalism in the context of black hole perturbation theory.

\section{Noncommutative perturbations of the Schwarzchild black hole} \label{secIII}

In this section, our investigation delves into the noncommutative deformation of black hole perturbations.
The metric $\oo{g}_{\mu \nu}$ characterizes the background spacetime of a black hole, and its perturbation is represented by the rank-2 symmetric tensor $h_{\mu \nu}$. The complete metric is therefore
\begin{equation}
	g_{\mu \nu} = \oo{g}_{\mu \nu} + h_{\mu \nu} .
\end{equation}
In our calculation, we adopt the Schwarzschild metric as the background metric $\oo{g}$:
\begin{equation}
   ds^2 = \oo g_{\mu \nu} dx^{\mu} dx^{\nu} = -\left(1- \frac{R}{r}\right) dt^2 + \left(1- \frac{R}{r}\right)^{-1} dr^2 + r^2\left(d\theta^2 + \sin^2 \theta d\varphi^2\right), 
\end{equation}
where $R = 2M$ denotes the horizon radius, and $M$ is the mass of the  black hole.  \\

Conventionally, the perturbation of a black hole in commutative theory is studied by imposing that the Ricci tensor is zero, $R_{\mu \nu}=0$, up to first order in $h$, following from the linearized Einstein equations\footnote{The components of any covariant tensor $\mathcal{\tau}$ are defined as $\mathcal{\tau}_{\mu_1 \mu_2 \cdots \mu_n}=\mathcal{\tau} (\partial _{\mu _1}, \partial _{\mu _2}, \cdots \partial _{\mu _n})$.}. Extending the notion of Einstein manifolds to noncommutative gravity requires careful consideration. 
Although one might intuitively consider $ \hat{R}  _{\mu \nu}=0$ as a straightforward extension by merely accounting for the correct commutative limit, the relation $\hat R_{\nu \mu}=0$, in contrast to the commutative case, will not be satisfied automatically here. This is because the definition of $\hat R_{\mu \nu}$ does not guarantee the symmetry  property and therefore imposing $\hat{R}_{\mu \nu} = 0$ leads to contradictions. To navigate this challenge, a recently proposed approach involves postulating a noncommutative Einstein manifold \cite{Herceg:2023zlk, Herceg:2023pmc} as
\begin{equation} \label{NCR}
    \hat{{\rm R}}_{\mu\nu}=0.
\end{equation}
Here, $\hat{{\rm R}}_{\mu\nu}$ is the $\mathcal{R}$-symmetrized Ricci tensor, defined as
\begin{equation}  
\label{NCRsymm}
	\hat{{\rm R}}_{\mu\nu}\equiv\frac{1}{2}\left\langle dx^{\alpha}, \hat{R} (\partial_{\alpha}, \partial_{\mu}, \partial_{\nu})+\hat{R} (\partial_{\alpha}, \bar{R}^{A}(\partial_\nu), \bar{R}_{A}(\partial_\mu) )\right\rangle_\star.
\end{equation}
Note that this proposal, first presented in \cite{Herceg:2023pmc, Herceg:2023zlk}, differs from the one found in \cite{Aschieri:2005zs, Aschieri:2009qh, Aschieri:2005yw, Schenkel:2011biz}. \\

As mentioned in the preceding section, we are in pursuit of noncommutative corrections to gravitational perturbations induced by the Moyal-Weyl twist \eqref{Moyal-Weyl}. For this type of deformation, $\mathcal{R}$-symmetrization reduces to the usual symmetrization, i.e.
\begin{equation}
	\hat{{\rm R}}_{\mu\nu}=\hat{R} _{(\mu\nu)} \equiv \frac{1}{2} (\hat{R} _{\mu\nu} + \hat{R} _{\nu\mu}).
\end{equation}

In addition, we consider an Abelian twist, a form of the Moyal-Weyl twist where globally defined vector fields $\{ X_\mu \}$ generating the twist commute. This twist has the form
\begin{equation} \label{abelian}
	\mathcal{F}=\exp \left( -i \Theta ^{\mu \nu} X _\mu \otimes X _\nu  \right).
\end{equation}
It is important to keep in mind that the selection of a specific twist is a matter of quantum-gravity phenomenology and ideally should be constrained by experiments. In the absence of such empirical inputs, one must rely on symmetry arguments to make a physically meaningful choice. The simplest approach in this regard might involve constructing a twist solely from a basis composed of Killing vectors $K^\mu$ of the background $\oo g_{\mu \nu}$. However, in the unperturbed background, it becomes apparent that such a choice would not yield a nontrivial noncommutative contribution. This is because any solution to the commutative Einstein equation is also a solution to the NC Einstein equation in cases where the twist is constructed from the Killing vectors of the nonperturbed background metric since $\pounds_K \oo g = 0$. Similar arguments apply to the choice of a semi-Killing twist, constructed using a Killing vector $K^\mu$ and an arbitrary basis vector $V^\nu$. For more details, see  \cite{Aschieri:2009qh, Herceg:2023pmc, Herceg:2023zlk, Schenkel:2011biz}.
In the realm of black hole perturbation studies, our focus is on the perturbed background black hole spacetime described by the metric $\oo{g}_{\mu \nu} + h_{\mu \nu}$. In this scenario, opting for the mere Killing twist seems to result in non-vanishing NC corrections that are quadratic in the metric perturbation $h$. 
This comes from the fact that $\pounds_{K} \oo g = 0$ and $\pounds_{K} h \neq 0$, resulting in only a quadratic leading-order term in noncommutative geometric quantities.  
In other words, if we want to respect the full symmetry of the background $\oo g$, then the NC effects are inherently nonlinear in perturbation, and we are forced to study quadratic commutative corrections, which would go beyond the linearized metric perturbation theory. \\

However, the leading linearized noncommutative metric perturbation term persists if we opt for a semi-pseudo-Killing twist constructed from the Killing field $K^\mu$ of the background $\oo g$ and an arbitrary vector $X^\nu$. We consider a semi-pseudo-Killing twist of the form:
\begin{equation} \label{semipseudo}
    \mathcal{F} = e^{-i \frac{a}{2} \big(K \otimes X - X \otimes K \big)}.
\end{equation}
%where we assume that the components of the antisymmetric matrix $\Theta ^{\mu \nu}$ are simply the Levi-Civita symbol. $
In the spherical coordinate basis, two Killing fields, namely $\partial_t$ and $\partial_\varphi$ are implemented as partial derivatives which makes them suitable for constructing the field $K$:
\begin{equation} \label{kxdef}
    K = \alpha \partial_t + \beta \partial_\varphi, \qquad X = \partial_r, \qquad \alpha, \beta \in \mathbb{R}.
\end{equation}
This choice produces the following commutation relations between the coordinates:
\begin{align*}
	[t\stackrel{\star}{,} r] &= i a \alpha,\\
	[\varphi \stackrel{\star}{,}r] &= i a \beta.
\end{align*}
Both commutators display quantization of the radial coordinate as this is necessary in order to have noncommutative corrections to perturbations at the linear level.
Furthermore, we introduce the parameter $\lambda$ as an eigenvalue of the Killing field's action on the perturbation:
\begin{align}
	h_{\mu \nu} \propto e^{i m \varphi}e^{- i \omega t} \implies \pounds_K h_{\mu \nu} = i \lambda \: h_{\mu \nu}, \\
	\text{   for } \quad K = \alpha \partial_t + \beta \partial_\varphi, \quad
	\lambda = -\alpha \omega + \beta m. \label{alfabeta}
\end{align}
In the calculation of the noncommutative entities, we can use the following linearized $\star$-product operation following \eqref{semipseudo} and \eqref{genericstar}:
\begin{equation}
	h \star k = h k + \frac{i}{2} a \big( K(h) X(k)-X(h) K(k) \big) + \mathcal{O}(a^2),
\end{equation}
where $h, k \in C_\star^\infty(\mathcal{M})$ and $V(f) := \pounds_V f$ for generic vector field $V^\mu$ and function $f$.\\

The spherically symmetric nature of the Schwarzschild background allows us to decompose the perturbations $h_{\mu \nu}$ into tensor spherical harmonics $Y^n_{\ell m}(\theta, \varphi)$, which can all be expressed in terms of $Y_{\ell m}(\theta, \varphi)$ and their derivatives with respect to $\theta$ and $\varphi$.
Schwarzschild background allows us to decompose perturbations $h_{\mu \nu}$ into scalar, vectorial, and tensorial spherical harmonics, which are certain derivatives with respect to $\theta$ and $\varphi$ of $Y_{\ell m}(\theta, \varphi)$. 
These perturbations manifest as axial and polar modes, each displaying distinct behaviors under the parity transformation $\vec{r} \rightarrow -\vec{r}$. Even-parity or polar modes, akin to scalar spherical harmonics $Y_{\ell m}(\theta, \varphi)$, transform as $(-1)^\ell$, while axial or odd-parity modes transform as $(-1)^{\ell+1}$. In the linear order, these modes are decoupled, facilitating separate treatment. \\

We will calculate the equations of motion up to linear order in both perturbation $h_{\mu \nu}$ and the NC deformation $a$. 
Due to the semi-psuedo Killing form of the twist, the perturbation and deformation parts are inherently coupled, appearing in pairs.

\subsection{Axial perturbations and noncommutative Regge-Wheeler potential} \label{sec3A}

Here we provide a brief overview of noncommutative axial perturbations \cite{Herceg:2023lmt, Herceg:2023pmc, Herceg:2023zlk}. 
The idea is to mimic the procedure of Regge and Wheeler \cite{Regge:1957td} in the NC setting by deforming the geometric quantities using methods outlined in the previous section. \\

Generally, axial perturbations are characterized by three families of radial functions $h_0^{\ell m}(r)$, $h_1^{\ell m}(r)$, and $h_2^{\ell m}(r)$. However, in the Regge-Wheeler gauge (see Appendix A in \cite{Herceg:2023pmc} for details), one can choose $h_2^{\ell m}=0$, resulting in only four non-zero components of the perturbation metric \cite{Langlois:2021xzq}:
\begin{equation}\label{ansatz}
\begin{split}
	h_{t \theta}&=\frac{1}{\sin \theta} \sum_{\ell, m} h_0^{\ell m} \partial_{\phi} Y_{\ell m}(\theta, \phi)e^{-i \omega t},\\
	 h_{t \phi}&=-\sin \theta \sum_{\ell, m} h_0^{\ell m} \partial_\theta Y_{\ell m}(\theta, \phi)e^{-i \omega t},\\
	h_{r \theta}&=\frac{1}{\sin \theta} \sum_{\ell, m} h_1^{\ell m} \partial_{\phi} Y_{\ell m}(\theta, \phi) e^{-i \omega t},\\
	 h_{r \phi}&=-\sin \theta \sum_{\ell, m} h_1^{\ell m} \partial_\theta Y_{\ell m}(\theta, \phi) e^{-i \omega t}.
\end{split}
\end{equation}
Similar to the commutative case \cite{Regge:1957td, Edelstein:1970sk}, we obtain only 7 non-zero components of the equation \eqref{NCR}. These components are in separable form, resulting in only 3 distinguishable (mutually distinct) radial components. \\

\begin{figure}[t]
\centering
	\subfigure[]{\includegraphics[scale=0.7]{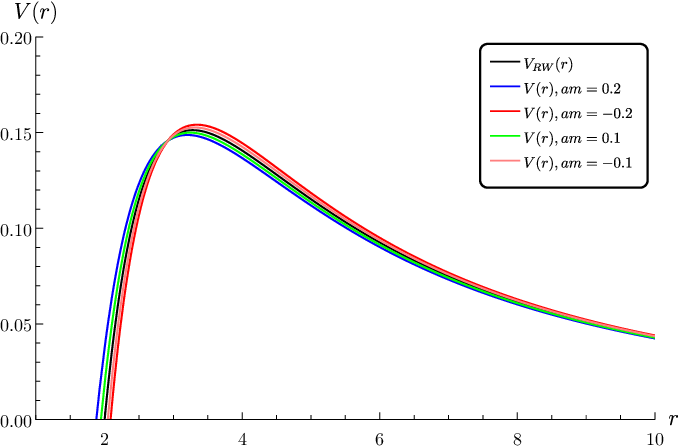}\label{figRW1}} 
\qquad
	\subfigure[]{\includegraphics[scale=0.7]{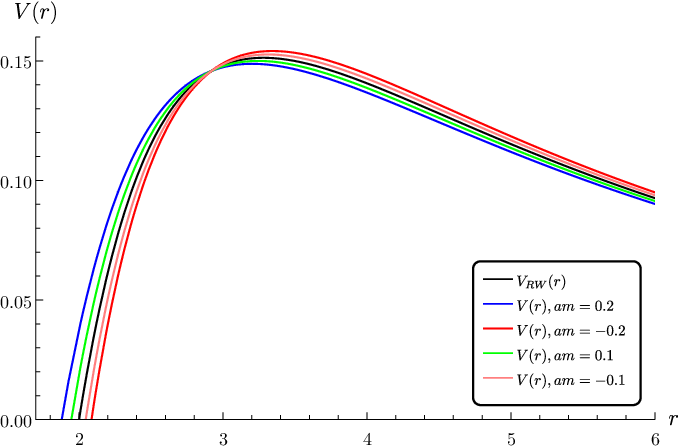}\label{figRW2}} 
	\caption{The figure shows the behavior of the potential for $R=2$ ($M=1$) and $\ell=2$. Both noncommutative and commutative cases exhibit similar behavior outside the horizon, particularly around the peak at $r_0\cong 1.5 R$ and for larger $r$. The right panel highlights the behavior of the potential near the peak.} \label{proba}
\end{figure}

\begin{figure}[t]
\centering
\includegraphics[scale=1.0]{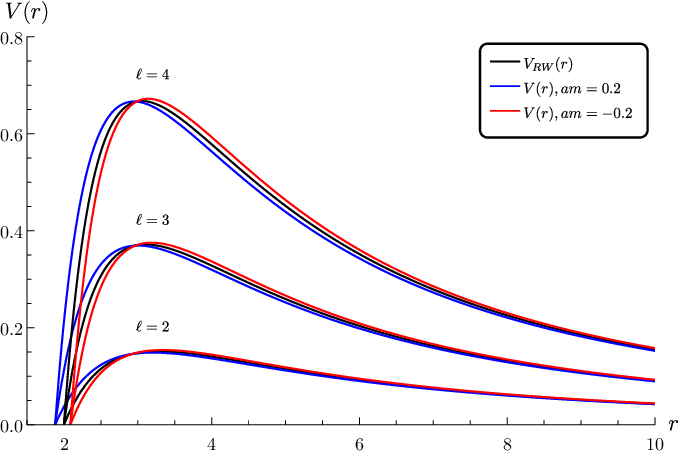}
	\caption{Plot of the axial NC potential for $R=2$ ($M=1$) and $\ell=2,3,4$.} \label{RWL}
\end{figure}

From $\hat{{\rm R}}_{r\phi}=0$ we get
\begin{equation} \label{RWeq1}
\begin{split}
       &  4 i r^4 (r - R)\omega h_0 + 2 r^2 (r - R) \big(r^3 \omega^2 - (r - R)(\ell(\ell + 1) -2)\big)h_1 - 2 i \omega r^5(r - R) h_0' \\
 & + \lambda a \Big[ 2 i r^3 \omega (r - 2R) h_0 +
\big((2\ell(\ell+1)+12) r (r- R)^2 \\
&- 9(r - R)^2 R - r^4 R \omega^2\big)h_1 + i
r^4 R \omega h_0' + 2 r (r - R)^3 h_1' \Big] =0,
\end{split} 
\end{equation}
from $\hat{{\rm R}}_{t\phi}=0$ we get
\begin{equation} \label{RWeq2}
\begin{split}
          &2r (2R - \ell(\ell+1)r) h_0 +
4 i r^2 \omega (r - R) h_1   + 2 r^3(r - R)(i \omega h_1' + h_0'')\\
  &+ \lambda a \Big[  (2 \ell (\ell+1) r + R)h_0   + i r \omega (4 r - 3 R) h_1 + r (4r - 5 R) h_0' + r^2 R ( i\omega h_1' + h_0'') \Big] = 0,
\end{split} 
\end{equation}
and from $\hat{{\rm R}}_{\theta\phi}=0$ we get
\begin{equation} \label{RWeq3}
\begin{split}
     & \frac{i r^3 \omega}{r - R}h_0 + R h_1 + r(r
- R)h_1' - \lambda a \Big[ \frac{i r^2 R \omega}{2(r - R)^2}h_0 - 3
\frac{r - R}{r} h_1 - \frac{1}{2}R h_1'\Big]=0.
\end{split}
\end{equation}
The radial parts of the linearized equations $\hat{{\rm R}}_{r \theta}=0$, $\hat{{\rm R}}_{\theta \theta}=0$, $\hat{{\rm R}}_{\varphi \varphi}=0$, and $\hat{{\rm R}}_{t \theta}=0$ are identical to the above equations. Among the 3 equations (\ref{RWeq1}-\ref{RWeq3}), only 2 are mutually independent. 
We are therefore in position to pick any two out of these three equations to find the solution of the whole system, the easiest choice being first-order equations \eqref{RWeq1} and \eqref{RWeq3}. Combining these equations eliminates $h_0$ and yields a single second-order differential equation for $h_1$. 
\begin{equation}\begin{split}
	&r (r-R) \Big( \ell(\ell + 1) r (R-r) + 2r^2 - 6rR + 5R^2 + \omega^2 r^4 \Big) h_1 +r^2(r - R)^2\Big( (5R - 2r)h_1' + r(r - R)h_1''\Big) \\
 &+ \lambda a \bigg[\Big(\ell(\ell + 1)r(r - R)^2 - 6r^3 + \frac{R}{2}(49
r^2 - 64 r R + 26 R^2 - \omega^2 r^4) \Big) h_1
 + r(r - R)^2 \Big( 3 ( r
- 2R) h_1' + \frac{1}{2}r R h_1''\Big) \bigg]=0. 
\end{split}\end{equation}
This can be re-expressed as a Schr{\"o}dinger-like equation with a field redefinition
\begin{equation}
h_1(r) = \frac{r^2}{r - R}\Big[ 1 + \frac{\lambda a}{2} \Big( \frac{3}{r} - \frac{1}{r - R} + \frac{1}{R} \log \frac{r}{r - R} \Big) \Big]\psi(r)
\end{equation}
and a noncommutative tortoise variable transformation
\begin{equation} \label{RWtortoise}
\hat{r}_* = r + R \log \frac{r - R}{R} + \frac{\lambda a}{2} \frac{R}{r - R}.
\end{equation}
Expressions for the tortoise coordinate and field redefinition come from two algebraic conditions that we have to impose in order to fix the Schr\"odinger form of the equation.
Detailed derivation is given in Appendix C of \cite{Herceg:2023pmc}. 
Finally, we arrive at the differential equation,
\begin{equation}\label{Rschrodinger}
	\frac{d^2 \psi}{d{\hat r}_*^2} + \Big( \omega^2 - V(r) \Big) \psi = 0,
\end{equation}
where $V(r)=V_{RW}+V_{NC}$ represents the noncommutative Regge-Wheeler potential,
\begin{equation}\begin{split}\label{RWpotential}
V(r) &= \frac{(r - R)\big(\ell (\ell + 1)r - 3R\big)}{r^4} + \lambda a \frac{\ell(\ell + 1)(3R - 2r)r + R(5r - 8R)}{2 r^5}.
\end{split}
\end{equation}
In the limit $a\to 0$, only $V_{RW}$ remains, representing the commutative Regge-Wheeler potential \cite{Regge:1957td, Edelstein:1970sk}, while $V_{NC}$ denotes the noncommutative correction to the Regge-Wheeler potential. It is noteworthy that in case $\alpha = 0, \beta = 1$ in \eqref{alfabeta}, $V_{NC}$ is dependent on $\ell$ and $m$, leading to a Zeeman-like splitting as depicted in Figure \ref{proba}. 
From the figure, it is evident that two crucial aspects  of the potential, namely the peak and the zero, are influenced by the noncommutative structure of spacetime. This behavior of the NC potential persists for higher $\ell$ modes as can be seen from Figure \ref{RWL}. \\

In the commutative case, the potential vanishes precisely at the event horizon $r=R$. However, in the noncommutative case, the potential vanishes slightly either inside or outside the horizon, depending on the sign of $\lambda a$, i.e., $r=R \pm \frac{\lambda a}{2} + \mathcal{O}(a^2)$. 
This implies that each mode $m$ effectively 'perceives' a somewhat different position of the horizon. Physically, this characteristic can be understood as the introduction of a level of fuzziness to the event horizon in particle scattering. Specifically, particle scattering in the potential \eqref{RWpotential} can be described as a summation of various modes $\psi_{n \ell m}$, i.e., $\Psi=\sum \psi_{n \ell m}e^{-i\omega_n t}$. 
Since each mode experiences the position of the event horizon differently, the collective wave packet may perceive a slightly shifted event horizon instead of a sharply defined one.  \\

The second aspect altered by noncommutativity is the location, width and height of the peak of the potential. From an astrophysical perspective, this is intriguing as it will modify the photon sphere radius and the shadow of the black hole. We shall see in the next section that the noncommutative polar potential exhibits the same properties.

\subsection{Polar perturbations and noncommutative Zerilli potential} \label{sec3B}

In this subsection we follow the derivation of Zerilli \cite{Zerilli:1970se} to obtain NC corrections to the polar perturbations. The procedure outlined there can be adapted to the NC setting by deforming the geometry according to the rules of the previous section. Similar to the axial case, the solution of the NC vacuum Einstein equation \eqref{NCR} in the end turns out to be completely governed by the NC Zerilli potential in the Schr\"odinger equation. \\

Polar perturbations of the metric are characterized by seven families of functions. However, by exploiting the invariance of the theory under spacetime diffeomorphisms we can adopt a specific gauge, namely the Zerilli gauge (refer to Appendix \ref{Zerilli_gauge_app} for details). In this gauge, the polar perturbations are parameterized by four families of radial functions: $H_{0}^{\ell m}$, $H_{1}^{\ell m}$, $H_{2}^{\ell m}$, and $K^{\ell m}$ as described below:
\begin{eqnarray}
\label{eq:even-pert2}
&&h_{tt} = f(r)\sum_{\ell, m} H_{0}^{\ell m}(r) Y_{\ell m}(\theta,\varphi)e^{-i \omega t},  \quad
h_{tr} = \sum_{\ell, m} H_{1}^{\ell m}(r) Y_{\ell m}(\theta,\varphi)e^{-i \omega t}, \\
&&h_{rr} = \frac{1}{f(r)} \sum_{\ell, m} H_{2}^{\ell m}(r) Y_{\ell m}(\theta,\varphi)e^{-i \omega t},  \quad 
h_{ab} = \sum_{\ell, m} K^{\ell m}(r) \oo g_{ab} Y_{\ell m}(\theta,\varphi)e^{-i \omega t},
\end{eqnarray}
where $f(r)=1-R/r$. Here, the indices $a$ and $b$ represent angular coordinates $\theta$ and $\varphi$.
For this metric we find the $\star$-inverse metric, connection, curvature and Ricci tensor as defined in the previous section.
Finally, the linearized Einstein equation obtained from the $\mathcal{R}$-symmetrized Ricci tensor \eqref{NCR} gives rise to 10 coupled ordinary differential equations in $r$ due to the separability of the angular parts. 
Out of these 10 equations, only 7 are distinguishable.  \\

After separating the angular part, the $\hat{{\rm R}}_{\theta\varphi} = 0$ equation reduces to the following algebraic relation,
\begin{equation} \label{constraint1}
    H_0-H_2 +   \lambda a \left[\frac{R }{2 r(r-R)}H_0+\frac{R }{2 r(r- R)} H_2\right]=0. 
\end{equation}
This equation can be solved for $H_2$ in terms of $H_0$, allowing for the elimination of $H_2$ in the remaining equations,
\begin{equation}
    H_2=H_0 - \lambda a \frac{ R }{r (R-r)} H_0.
\end{equation}
In the commutative limit, this solution simplifies to $H_2 = H_0$. The remaining 6 equations involve only three unknown functions $K$, $H_0$, and $H_1$. \\

Equation $\hat{{\rm R}}_{tt} = 0$ reduces to

\begin{equation}
\label{eq:full-eom-schwarzschild1}
%\left\{
\begin{aligned}
%&\begin{multlined}
& \frac{2 (r-R) }{r^2} H_0' +\frac{(r-R)^2 }{r^2}H_0''+ \left(\frac{\ell (\ell+1) (R-r)}{r^3}-\omega ^2\right)H_0+\frac{   2 i \omega (r-R)}{r} H_1' \\
&  +\frac{   i \omega (4 r-3 R)}{r^2}H_1 +\frac{R (R-r) }{r^3}K'-2 \omega ^2 K \\
	&-\lambda a \Biggl[\frac{(r-R)}{r^3} H_0'+\frac{R (r-R) }{2 r^3}H_0''+ \left(\frac{(r-R) (\ell (\ell+1) r+R)}{r^5}+\frac{R \omega ^2}{2 r(r- R)}\right) H_0\Biggr. \\
	&\Biggl.  +\frac{i  \omega R  }{r^2}H_1'+\frac{ i\omega  (2 r-R)}{r^3} H_1 +\frac{R (4 r-5 R) }{2 r^4}K'+\frac{ \left(2 r^4 \omega ^2+R (r-R)\right)}{r^5}K \Biggr]=0.
%\end{multlined} \\
\end{aligned}
%\right.
\end{equation}

The $\hat{{\rm R}}_{tr} = 0$ equation reduces to

\begin{equation}
\label{eq:full-eom-schwarzschild2}
%\left\{
\begin{aligned}
%&\begin{multlined}
 & -\frac{H_0}{r}-\frac{i \ell (\ell+1) }{2 r^2 \omega } H_1 +K'+\frac{1}{2}  \left(\frac{1}{R-r}+\frac{3}{r}\right) K \\
& + \lambda a \left[\frac{ \left(2 r^2-6 r R+5 R^2\right)}{4 r^2 (r-R)^2}H_0 +\frac{i R (R-2 r)}{4 r^3 \omega  (r-R)}  H_1' +\frac{i}{4 r^4 \omega }  \left(\frac{R^2 (3 R-4 r)}{(r-R)^2}-2 \ell (\ell+1) r\right)H_1 \right. \\
&\left. +\frac{K'}{r} +\frac{ \left(4 r^2-12 r R+7 R^2\right)}{4 r^2 (r-R)^2} K \right]=0.
%&\begin{multlined} 
\end{aligned}
%\right.
\end{equation}

The $\hat{{\rm R}}_{rr} = 0$ equation reduces to

\begin{equation}
\label{eq:full-eom-schwarzschild3}
%\left\{
\begin{aligned}
%&\begin{multlined}
&  \frac{2 (r-R) }{r^2}H_0'+\frac{(r-R)^2 }{r^2}H_0''+ \left(\frac{\ell (\ell+1) (r-R)}{r^3}-\omega ^2\right)H_0+\frac{   2 i \omega (r-R)}{r} H_1'  \\
& +\frac{ i \omega  R }{r^2} H_1-\frac{(4 r-3 R) (r-R) }{r^3} K' -\frac{2 (r-R)^2 }{r^2}K''\\
&-\lambda a \left[\frac{\left(r^2-3 r R+R^2\right) }{r^4}H_0'+\frac{R (R-r) }{2 r^3} H_0'' \right.\\
&\left. +\frac{ 2 \ell (\ell+1) r^2 (R-r)+3 \omega ^2  r^4 R+2 R  (3 r-2 R)^2}{2 r^5 (r-R)} H_0 -\frac{i  \omega R}{r^2} H_1' +\frac{ i \omega  R }{r^3} H_1 \right. \\
&\left. +\frac{\left(8 r^2-12 r R+5 R^2\right) }{2 r^4} K' +\frac{2 (r-R)^2}{r^3}K''+\frac{R (r-R)}{r^5} K  \right]=0.
%&\begin{multlined}
\end{aligned}
%\right.
\end{equation}

The $\hat{{\rm R}}_{t\theta} = 0$ equation reduces to

\begin{equation}
\label{eq:full-eom-schwarzschild4}
%\left\{
\begin{aligned}
%&\begin{multlined}
& i \omega  H_0+\left(1-\frac{R}{r}\right) H_1'+\frac{R }{r^2}H_1+i \omega  K\\
& + \lambda a \left[\frac{i  \omega R }{2 r(r- R)} H_0 -\frac{R }{2 r^2}H_1' -\frac{ \left(4 r^2-9 r R+6 R^2\right)}{2 r^3 (r-R)} H_1 +\frac{i \omega}{r} K\right]=0.
%\end{multlined} \\
\end{aligned}
%\right.
\end{equation}

The $\hat{{\rm R}}_{r \theta} = 0$ equation reduces to

\begin{equation}
\label{eq:full-eom-schwarzschild5}
%\left\{
\begin{aligned}
%&\begin{multlined}
&\left(1-\frac{R}{r}\right) H_0'+\frac{R }{r^2}H_0+i \omega  H_1+\left(\frac{R}{r}-1\right) K'\\
& + \lambda a \left[\frac{R }{2 r^2}H_0'-\frac{ \left(4 r^2-9 r R+3 R^2\right)}{2 r^3 (r-R)}H_0+\frac{i \omega  R }{2 r(r-R)} H_1 -\frac{(r-R)}{r^2} K'-\frac{2  (R-r)}{r^3}K\right]=0.
\end{aligned}
%\right.
\end{equation}

The $\hat{{\rm R}}_{\theta \theta} = 0$ equation reduces to

\begin{equation}
\label{eq:full-eom-schwarzschild6}
%\left\{
\begin{aligned}
%&\begin{multlined}
%&\begin{multlined} 
 & 2( R- r) H_0'-2 H_0- 2 i \omega r H_1+(4 r-3 R) K'+r (r-R) K''+ \left(-\ell (\ell+1)+\frac{r^3 \omega ^2}{r-R}+2\right) K \\
& -\lambda a \left[\frac{(r+R) }{r}H_0'+ \left(\frac{3 R-4 r}{r^2}+\frac{2}{r-R}\right)H_0+\frac{ i \omega   (r+R)}{r-R} H_1 +\frac{1}{2} R K''+2 K' \right. \\
& \left. -\frac{1}{2 r^2} \left(-2 (\ell (\ell+1)+2) r+\frac{\omega ^2 r^4 R}{(r-R)^2}+6 R\right) K \right]=0.
%\end{multlined} \\
\end{aligned}
%\right.
\end{equation}

%In the above equations, terms enclosed within $[ \cdots ]$ represent commutative components, while those in $\{ \cdots \}$ denote leading order noncommutative corrections. 
	In the limit $a\rightarrow 0$, the linearized equations of motion derived from the condition $\hat{{\rm R}}_{\mu\nu}=0$ coincide with the equations obtained by Edelstein and Vishveshwara (Eq. $(9b)-(9g)$ in \cite{Edelstein:1970sk}).
The linearized equations $\hat{{\rm R}}_{t \varphi}=0$, $\hat{{\rm R}}_{r \varphi}=0$, and $\hat{{\rm R}}_{\varphi \varphi}=0$, in their radial part correspond identically to $\hat{{\rm R}}_{t \theta}=0$, $\hat{{\rm R}}_{r \theta}=0$ and $\hat{{\rm R}}_{\theta \theta}=0$, respectively. The combination
\begin{equation}
	\hat{{\rm R}}_{rr}-\left(1+\lambda a \frac{ R}{r (r-R)} \right)  \hat{{\rm R}}_{tt}-\left( -\frac{2 (r-R)}{r^3}-\lambda a \frac{(2 r-R)}{r^4}\right) \hat{{\rm R}}_{\theta\theta} = 0
\end{equation}
reduces the three second-order differential equations to a first-order relation

\begin{equation}
\begin{aligned}
	& -\frac{4 (r-R)^2 }{r^3}H_0'+\frac{2 (\ell (\ell+1)-2)  (r-R)}{r^3}H_0-\frac{8 i \omega   (r-R)}{r^2}H_1+\frac{2 \left(2 r^2-3 r R+R^2\right) K'}{r^3} \Biggr. \\
 & \Biggl. +\left(4 \omega ^2-\frac{2 (\ell (\ell+1)-2) (r-R)}{r^3}\right) K  \\
	&-   \lambda a \Biggl[ \frac{\left(6 r^2-6 r R+R^2\right) }{r^4}H_0'+\frac{ \left(-2 \ell (\ell+1) r^3+2 (\ell (\ell+1)+10) r^2 R-28 r R^2+9 R^3\right)}{r^5 (r-R)}H_0 \Biggr. \\
	&\Biggl. +\frac{i \omega   \left(4 r^2+2 r R-5 R^2\right)}{r^3 (r-R)}H_1+\frac{R (R-2 r) }{r^4}K'+ \left(\frac{(4 r-3 R) (\ell (\ell+1) r-2 R)}{r^5}-\frac{4 \omega ^2}{r-R}\right) K\Biggl] = 0.
\end{aligned}
\end{equation}
	The first-order terms in the above expression can be removed by employing the first-order equations \eqref{eq:full-eom-schwarzschild2}, \eqref{eq:full-eom-schwarzschild4} and \eqref{eq:full-eom-schwarzschild5}, resulting in an algebraic relation

\begin{equation} \label{algebraicrel}
    \begin{aligned}
        &  \left(-\ell (\ell+1)-\frac{3 R}{r}+2\right)H_0 + \left(2 i r \omega -\frac{i \ell (\ell+1) R}{2 r^2 \omega }\right) H_1  \\
       &  +\frac{\left(-2 (\ell (\ell+1)-2) r^2+2 (\ell (\ell+1)-3) r R+4 r^4 \omega ^2+3 R^2\right)}{2 r (R-r)}K\\
	    &-\lambda a \Biggl[  \left(\frac{\ell (\ell+1)-1}{r-R}+\frac{3 R}{2 r^2}-\frac{8}{r}\right) H_0+\frac{i  \left(-\ell (\ell+1) r (5 r-4 R) (r-R)+2 r^4 \omega ^2 (r-3 R)+R^3\right)}{2 r^4 \omega  (r-R)} H_1 \Biggr.\\
	    &\Biggl.+\frac{\left((18-4 \ell (\ell+1)) r^3+7 (\ell (\ell+1)-7) r^2 R+(47-3 \ell (\ell+1)) r R^2+4 r^5 \omega ^2-15 R^3\right)}{2 r^2 (r-R)^2} K \Biggr]=0 .
    \end{aligned}
\end{equation}
	The above derivation demonstrates that any of the second-order equations in \eqref{eq:full-eom-schwarzschild1}-\eqref{eq:full-eom-schwarzschild6} can be derived from the three first-order equations, provided the algebraic relationship \eqref{algebraicrel} holds. Therefore, it suffices to concentrate on the first-order equations to obtain the complete solution for the system of equations \eqref{eq:full-eom-schwarzschild1}-\eqref{eq:full-eom-schwarzschild6}. However, one can prove that one equation among the three is redundant -- the algebraic equation eliminates $H_0$ from the first-order equations, leaving three equations in two functions $K$ and $H_1$. The dependence of the three equations can be proved in a similar way to appendix B of \cite{Herceg:2023pmc}. \\

Solving the system therefore amounts to solving for $K$ and $H_1$ from \eqref{eq:full-eom-schwarzschild2} and \eqref{eq:full-eom-schwarzschild4}.
Rearranging these equations, while eliminating $H_0$ with the aid of algebraic relation, yields two first-order equations (with redefinition$\footnote{This redefinition is introduced to yield a system of equations that does not involve powers of $\omega$ higher than 2, corresponding to at most second-order equations in time.}\ L=H_1/\omega$):
\begin{equation}
    \begin{aligned} \label{coupled_ode}
        K' &= \left[\alpha _0 (r) + \alpha _2 (r) \omega ^2\right] K + \left[\beta _0 (r) + \beta _2 (r) \omega ^2\right] L, \\
        L' &= \left[\gamma _0 (r) + \gamma _2 (r) \omega ^2\right] K + \left[\delta _0 (r) + \delta _2 (r) \omega ^2\right] L,
    \end{aligned}
\end{equation}
where $\alpha (r)$, $\beta (r)$, $\gamma (r)$, and $\delta (r)$, with the abbreviation $\ell (\ell+1)=2\Lambda +2$ are given below:
\begin{equation}
    \begin{aligned}
        \alpha _0(r)= & \frac{R ((\Lambda -2) r+3 R)}{r (r-R) (2 \Lambda  r+3 R)} \\
	    &-\lambda a \Biggl[\frac{\left(4 (\Lambda -2) \Lambda  r^4+4 \left(4-5 \Lambda ^2\right) r^3 R+(\Lambda  (12 \Lambda -17)-56) r^2 R^2+15 (\Lambda +3) r R^3-9 R^4\right)}{2 r^2 (r-R)^2 (2 \Lambda  r+3 R)^2}\Bigg],\\
	    \alpha _2(r)= & -\frac{2 r^3}{(r-R) (2 \Lambda  r+3 R)}-\lambda a \Biggl[ \frac{2  r^2 \left((7-3 \Lambda ) r^2+(5 \Lambda -11) r R+9 R^2\right)}{(r-R)^2 (2 \Lambda  r+3 R)^2} \Biggr],\\
        \beta _0(r)=& \frac{2 i (\Lambda +1) (\Lambda  r+R)}{r^2 (2 \Lambda  r+3 R)}\\
	    &-\lambda a \Biggl[\frac{i \left(-10 \Lambda  (\Lambda +1) r^4+(\Lambda +1) (13 \Lambda -8) r^3 R+(13-5 (\Lambda -2) \Lambda ) r^2 R^2-(8 \Lambda +3) r R^3-3 R^4\right)}{r^4 (r-R) (2 \Lambda  r+3 R)^2}\Biggr],\\
	    \beta _2(r)=& \frac{2 i r}{2 \Lambda  r+3 R} -\lambda a \Biggl[\frac{i \left(2 (6 \Lambda -7) r^2+(31-16 \Lambda ) r R-27 R^2\right)}{(r-R) (2 \Lambda  r+3 R)^2} \Biggr],\\
	    \gamma _0 (r)= & -\frac{i r (8 \Lambda  r (r-R)+R (8 r-9 R))}{2 (r-R)^2 (2 \Lambda  r+3 R)}\\
	    &-\lambda a \Biggl[ \frac{i }{4 (r-R)^3 (2 \Lambda  r+3 R)^2}\left(32 \Lambda ^2 r^4+16 \Lambda  r^4-48 \Lambda ^2 r^3 R+96 \Lambda  r^3 R-32 r^3 R+16 \Lambda ^2 r^2 R^2 \right. \\
	    &\left. -188 \Lambda  r^2 R^2+166 r^2 R^2+72 \Lambda  r R^3-225 r R^3+81 R^4\right) \Biggr], \\
	    \gamma _2 (r)= & \frac{2 i r^5}{(r-R)^2 (2 \Lambda  r+3 R)}+\lambda a\Biggl[\frac{i r^4 \left(14 r^2+4 \Lambda  r R-13 r R+9 R^2\right)}{(r-R)^3 (2 \Lambda  r+3 R)^2} \Biggr], \\
        \delta _0 (r)= &  -\frac{R (3 \Lambda  r+r+3 R)}{r (r-R) (2 \Lambda  r+3 R)} \\
	    &-\lambda a \Biggl[ -\frac{1 }{2 \left(r^2 (r-R)^2 (2 \Lambda  r+3 R)^2\right)}\left( 36 \Lambda ^2 r^4+20 \Lambda  r^4-68 \Lambda ^2 r^3 R+32 \Lambda  r^3 R+16 r^3 R \right.\\
	    &\left.+32 \Lambda ^2 r^2 R^2-131 \Lambda  r^2 R^2+r^2 R^2+73 \Lambda  r R^3-66 r R^3+42 R^4\right) \Biggr],\\ 
	    \delta _2 (r)= & \frac{2 r^3}{(r-R) (2 \Lambda  r+3 R)}- \lambda a \Biggl[\frac{2 r^2 \left(3 \Lambda   r^2-7  r^2-5 \Lambda   r R+11  r R-9  R^2\right)}{(r-R)^2 (2 \Lambda  r+3 R)^2} \Biggr].
    \end{aligned}
\end{equation}

To cast the coupled differential equations \eqref{coupled_ode} in a form resembling the Schr{\"o}dinger equation, we introduce a field redefinition
\begin{equation} \label{transformation}
    \begin{aligned}
        K = \ &\hat{f}(r)\hat{K}+\hat{g}(r)\hat{L},\\
        L = \ &\hat{h}(r)\hat{K}+\hat{l}(r)\hat{L},\\
    \end{aligned}
\end{equation}
accompanied by a coordinate transformation $dr/d\hat r_*=\hat n(r)$. 
Our aim is to find the functions $\hat f, \hat g, \hat h$, $\hat l$ and $\hat n$ such that the following conditions hold:
\begin{equation} \label{requirement} 
    \begin{aligned}
	    \frac{d\hat{K}}{d\hat r_*}=\hat{L}, \qquad \frac{d\hat{L}}{d\hat r_*}=(V-\omega ^2)\hat{K},
    \end{aligned}
\end{equation}
which, when combined, form the Schr{\"o}dinger equation for $\hat K$,
\begin{equation} \label{Zschrodinger}
    \begin{aligned}
       \frac{d^2\hat{K}}{d\hat r^2_*}+(\omega ^2 -V)\hat{K}=0.
    \end{aligned}
\end{equation}
The system \eqref{coupled_ode} expressed in terms of $\hat K$ and $\hat L$ is 
\begin{equation} \label{coupled_ode2}
    \begin{aligned}
	    \frac{d \hat K}{d \hat r_*} =&\left[\hat \alpha _0 (r)+ \hat \alpha _2 (r) \omega ^2\right] \hat K+ \left[\hat \beta _0 (r)+\hat \beta _2 (r) \omega ^2\right] \hat L, \\
	    \frac{d \hat L}{d \hat r_*} =&\left[\hat \gamma _0 (r)+ \hat \gamma _2 (r) \omega ^2\right] \hat K+ \left[ \hat \delta _0 (r)+ \hat \delta _2 (r) \omega ^2\right] \hat L.
    \end{aligned}
\end{equation}
The requirement \eqref{requirement} imposes the following specific constraints on the coefficient functions:
\begin{equation} \label{trans-constraints}
    \begin{aligned}
      \hat \alpha _0 (r)= \hat \alpha _2 (r)= \hat \beta _2 (r)=\hat \delta _0 (r)=\hat \delta _2 (r) =0, \qquad
      \hat \beta _0 (r)=1, \qquad \hat \gamma _2 (r)=-1 .
    \end{aligned}
\end{equation}
Our requirement involves seven conditions for five unknowns: the coefficients $\hat{f}(r)$, $\hat{g}(r)$, $\hat{h}(r)$, $\hat{l}(r)$ and the tortoise coordinate $\hat n(r)$. Fortunately, the additional degrees of freedom coincide, allowing the system to admit a solution. Moreover, the solution reduces to the commutative result in the limit $a\rightarrow 0$. The exact functional forms of the coefficients of the transformation equation \eqref{transformation} are complicated and are provided in Appendix \ref{app-coeff}. The solutions we obtained have the form
\begin{equation} \label{eqfghl}
    \begin{aligned}
	    \hat{f}(r)=&\ f(r)- \, \lambda a \, \tilde f(r), \qquad \hat{g}(r)= g(r)- \, \lambda a \, \tilde g(r),\\
	    \hat{h}(r)=&\ h(r)- \, \lambda a \, \tilde h(r), \qquad \hat{l}(r) = l(r)- \, \lambda a \, \tilde l(r),
    \end{aligned}
\end{equation}
where $\{ f(r), g(r), h(r), l(r)\}$ are the commutative components and $\{\tilde f(r), \tilde g(r), \tilde h(r), \tilde l(r)\}$ denote the noncommutative corrections. The NC tortoise coordinate of the transformation is given by
\begin{equation} \label{Ztortoise}
	\hat r_* =\ r + R \log \frac{r - R}{R} - \lambda a \left(\frac{(2 \Lambda +7) R}{2 (2 \Lambda +3)
   (r-R)}-\frac{4 (\Lambda +3) \log
   \left(\frac{r}{R}-1\right)}{(2 \Lambda +3)^2}-\frac{9 \log
   \left(\frac{2 \Lambda  r}{R}+3\right)}{\Lambda  (2 \Lambda
	+3)^2}\right)
\end{equation}
where we used the abbreviation $\ell (\ell+1)=2\Lambda +2$. 
Similar to the axial case, the tortoise coordinate and field redefinition come from the two conditions \eqref{requirement} that we impose  in order to fix the Schr\"odinger form of the equation.
Note that the noncommutative part of the tortoise coordinate differs from that of axial perturbation (see Ref. \cite{Herceg:2023pmc}) since it depends on both $\ell$ and $m$. The resulting Schr{\"o}dinger-like form of the equation \eqref{Zschrodinger} features an effective potential expressed as $V=V_\text{Z}+\, V_\text{NC}$, where $V_\text{Z}$ denotes the commutative Zerilli potential, and $V_\text{NC}$ represents the noncommutative correction. The explicit forms of these potentials are
\begin{equation} \label{Zpotential}
    \begin{aligned}
        V_\text{Z}=&\ \frac{(r-R) \left(8 \Lambda ^2 (\Lambda +1) r^3+12 \Lambda ^2 r^2 R+18 \Lambda  r R^2+9 R^3\right)}{r^4 (2 \Lambda  r+3 R)^2},\\
        V_\text{NC}=&\ \frac{\lambda a}{4 r^5 (2 \Lambda  r+3 R)^3} \Big[32 \Lambda ^2 \left(2 \Lambda ^2+7\right) r^5-8 \Lambda ^2 (2 \Lambda  (6 \Lambda -13)+121) r^4 R  \\
        & \quad -12 \Lambda  (2 \Lambda  (15 \Lambda -58)+59) r^3 R^2-2 (\Lambda  (440 \Lambda -741)+162) r^2 R^3 -3 (316 \Lambda -207) r R^4-387 R^5 \Big].
    \end{aligned}
\end{equation}

\begin{figure}[t]
\centering
\subfigure[ref2][]{\includegraphics[scale=0.7]{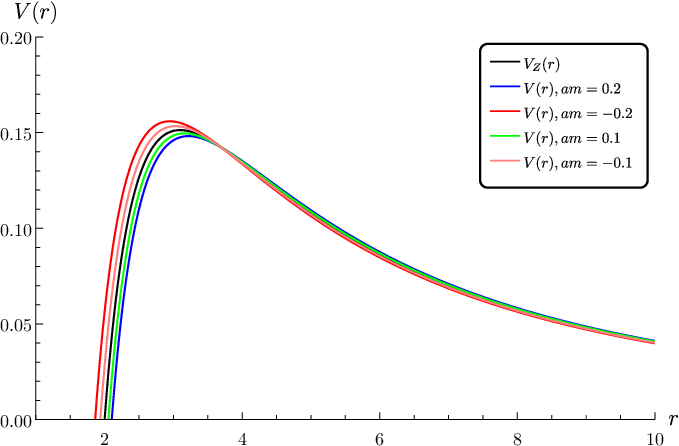}
\label{nczerilli1}}
\qquad
\subfigure[ref3][]{\includegraphics[scale=0.7]{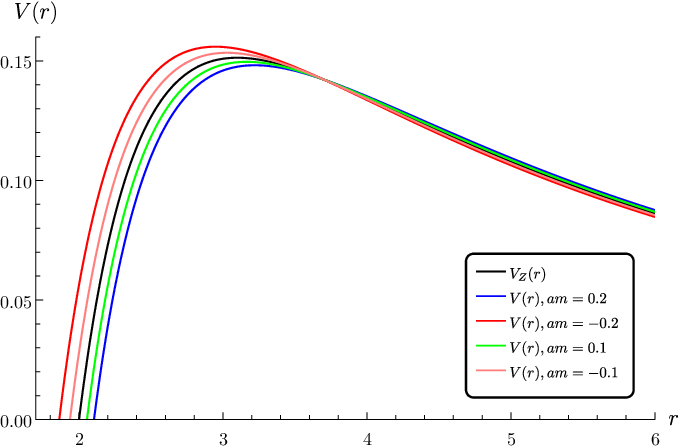}
\label{nczerilli2}}
\caption{Plot of the potential for \(R=2\) (\(M=1\)) and \(\ell=2\) for a wider range of \(r\) and \(V(r)\) is on the left. Near-horizon region is shown on the right.} \label{nczerilli}
\end{figure}
%
%
%%%%%%%%%%%%%%%%
%
The potential $V = V_Z + V_{NC}$ is illustrated in Figure \ref{nczerilli1}. Similar to the axial case, the noncommutative structure of spacetime influences two significant characteristics of the potential: its peak and zero. In the commutative scenario, the polar potential precisely vanishes at the horizon. However, in the noncommutative case, the polar potential either vanishes slightly inside or outside the horizon, depending on the sign of the correction term. This phenomenon, as discussed in the axial case, can be attributed to the fuzziness introduced by noncommutativity around the black hole horizon. The behavior of the polar potential around its peak is depicted in Figure \ref{nczerilli2}. Similar to the axial case, deviations observed around the peak of the potential, at approximately $r_0\approx 1.5 R$, can be interpreted as Zeeman-like splitting in case of pure $r-\varphi$ noncommutativity ($\alpha = 0, \beta = 1 \implies \lambda = m$).
Contrasting Figure \ref{figRW2} with \ref{nczerilli2}, we see that polar perturbations seem to alter the width of the potential by a larger margin. The same goes for the height, although the difference is less prominent. \\

\begin{figure}[t]
\centering
\includegraphics[scale=1.0]{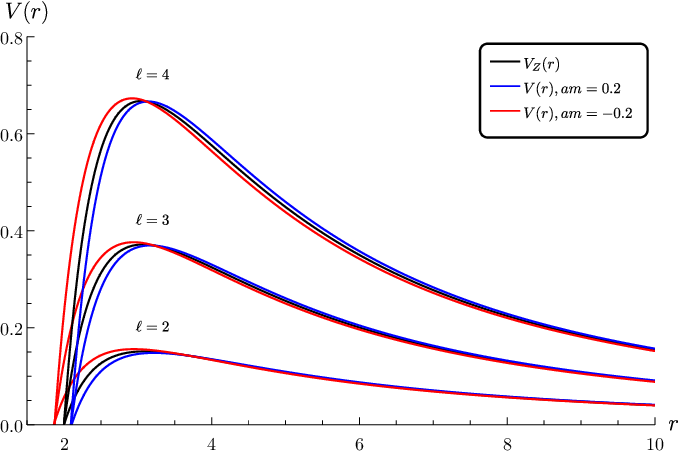}
	\caption{Plot of the polar NC potential for $R=2$ ($M=1$) and $\ell=2,3,4$.} \label{grafl}
\end{figure}

It is worth noting that a similar splitting of the potential and the corresponding QNM spectrum was observed in \cite{Ciric:2017rnf, DimitrijevicCiric:2019hqq} in the context of NC  charged scalar field probing a background of the Reissner-Nordstr\"om black hole.
This property reminiscent of Zeeman-like splitting, along with a distinct peak around $r_0\cong 1.5 R$ persists for higher multipoles as well, as illustrated in Figure \ref{grafl}. Additionally, Figure \ref{grafl} demonstrates that the shifting of the potential near the horizon follows a consistent trend across all modes.
Interestingly, zeros of the potential (and the position of the peak) are translated oppositely relative to the axial case. This is visible from the ordering of blue and red lines in Figures \ref{figRW2} and \ref{nczerilli2}.

\section{Noncommutative corrections to quasinormal modes} \label{secIV}

In this section, we will employ semi-analytical methods to compute the quasinormal frequencies of both noncommutative axial and polar perturbations. By analyzing the results of these calculations, we aim to assess the stability of the black hole spacetime under perturbation and how the presence of noncommutativity affects the QNM spectra.

\subsection{The WKB Method}
One of the most elegant methods for calculating quasinormal modes in black hole perturbation theory is the WKB (Jeffreys-Wentzel-Kramers-Brillouin) approximation. This method was initially introduced in the context of black hole scattering theory by Schutz and Will \cite{Schutz:1985km}. The key concept in this approach involves matching the solutions at spatial infinity and the black hole horizon using a Taylor expansion in the vicinity of the maximum of the potential barrier through the two turning points. Here, the Schr{\"o}dinger-like equation is reformulated into the form,
\begin{equation}\label{WKB_wave}
\frac{d^2\Psi}{dx^2} + Q(x)\Psi(x) = 0,
\end{equation}
where $x = r_*$ and $Q(x) = \omega^2 - V$. Since $-Q_{\text{max}}(x) \ll Q(\pm \infty)$, the solution near the peak of the potential can be approximated by the Taylor series,
\begin{equation} \label{WKB_Taylor}
Q(x) = Q_0 + \frac{1}{2}Q_0^{\prime \prime}(x - x_0)^2 + \mathcal{O}((x - x_0)^3),
\end{equation}
where $x_0$ is the location of the maximum with the function value $Q_0 = Q(x_0)$ and second derivative $Q_0^{\prime \prime}$ is with respect to $x$ at the peak. The $Q_0^{\prime}$ term does not appear in the series expansion as it vanishes at the maximum. Substituting \eqref{WKB_Taylor} into \eqref{WKB_wave} results in a parabolic cylinder differential equation, known as the Weber equation. At first order, the solution to this equation reads (in terms of the effective potential),
\begin{equation}
\omega^2 = V_0 - i\sqrt{-2V_0^{\prime \prime}}\left(n + \frac{1}{2}\right).
\end{equation}
Here, $n$ denotes the harmonic or overtone number. While the analytical expression for the quasinormal mode frequency is derived, obtaining an analytical expression for $x_0$ or the tortoise coordinate at the peak of the potential is not always feasible. Thus, the WKB approach is considered semi-analytical in this regard. \\

To enhance precision, the WKB method has been extended to the third order by Iyer and Will \cite{Iyer:1986np}. The third-order formula for QNM frequencies is given by
\begin{equation}
\omega^2=\left[V_0+\sqrt{-2 V_0^{\prime \prime}} \Lambda_1\right]-i\nu\sqrt{-2 V_0^{\prime \prime}}[1+\Lambda_2],
\label{204}
\end{equation}
where
\begin{widetext}
\begin{subequations}
\begin{align}
\Lambda_1&=\frac{1}{\sqrt{-2 V_0^{\prime \prime}}}\left[\frac{1}{8}\left(\frac{V_0^{(4)}}{V_0^{\prime \prime}}\right)\left(\frac{1}{4}+\nu^2\right)-\frac{1}{288}\left(\frac{V_0^{(3)}}{V_0^{\prime \prime}}\right)^2\left(7+60 \nu^2\right)\right],\label{3order1}\\
\Lambda_2 &=-\frac{1}{2 V_0^{\prime \prime}}\left[\frac{5}{6912}\left(\frac{V_0^{(3)}}{V_0^{\prime \prime}}\right)^4\left(77+188 \nu^2\right)-\frac{1}{384} \frac{V_0^{(3) 2} V_0^{(4)}}{V_0^{\prime \prime 3}}\left(51+100 \nu^2\right)\right.\label{3order2} \\ 
&+\frac{1}{2304}\left(\frac{V_0^{(4)}}{V_0^{\prime \prime}}\right)^2\left(67+68 \nu^2\right)+\frac{1}{288} \frac{V_0^{(3)} V_0^{(5)}}{V_0^{\prime \prime 2}}\left(19+28 \nu^2\right)\left.-\frac{1}{288} \frac{V_0^{(6)}}{V_0^{\prime \prime}}\left(5+4 \nu^2\right)\right] \notag ,
\end{align}
\end{subequations}
\end{widetext}
with $\nu = n + 1/2$, and $V_0^{(j)}$ denoting the $j^{th}$ derivative of the potential evaluated at its peak in tortoise coordinate. Subsequently, the method was extended to the sixth order by Konoplya \cite{Konoplya:2003ii}, resulting in even greater accuracy. The expression for QNM frequencies is given by,
\begin{equation}
    \frac{i(\omega ^2-V_0 )}{\sqrt{-2V_0^{\prime \prime}}}-\sum _{i=2}^{6} \tilde \Lambda _i=n+\frac{1}{2}.
\end{equation}
The higher-order corrections $\tilde \Lambda _i$ involve complicated functions of derivatives of potential up to $V_0^{(12)}$, with explicit expressions available in \cite{Konoplya:2003ii}. Furthermore, the WKB method has been developed up to the thirteenth order by Matyjasek and Opala \cite{Matyjasek:2017psv}. However, it is noteworthy that higher-order WKB does not always lead to superior approximations due to the fact that WKB expansion is an asymptotic series \cite{Konoplya:2019hlu}. The optimal order for minimizing the error in QNM frequency calculation depends on the specific effective potential. The error is estimated by comparing the solution of two consecutive orders using the formula \cite{Konoplya:2019hlu}
\begin{equation}
    \Delta _k=\frac{| \omega _{k+1}-\omega _{k-1}|}{2},
\end{equation}
where for the sake of comparison, one usually takes fundamental mode  frequency.
To obtain the numerical values for the frequencies for the noncommutative potentials \eqref{RWpotential} and \eqref{Zpotential}, we must specify the exact kind of twist.
This amounts to specifying constants $\alpha, \beta$ and $a$ in the equation \eqref{kxdef}. 
WKB analysis turns out to be simpler for the twist $\alpha = 0, \beta = 1$, to which we stick from now on.
In that case, the only nontrivial commutator is the one between the radial and angular coordinate, $[\varphi \stackrel{\star}{,}r] = i a$ and all occurences of $\lambda a$ in the previous section can be replaced by $a m$ as per \eqref{alfabeta}. 
This noncommutative space was analyzed in \cite{Herceg:2023zlk}, where the NC-corrected axial potential was presented, and the axial QNM frequencies were calculated using the third-order WKB method. In that work, it was referred to as $\qbar$-space, where $\qbar$ plays the role of noncommutativity parameter $a$ in the case of pure $r-\varphi$ twist.
\\

\begin{figure}[t]
\centering
\includegraphics[scale=0.80]{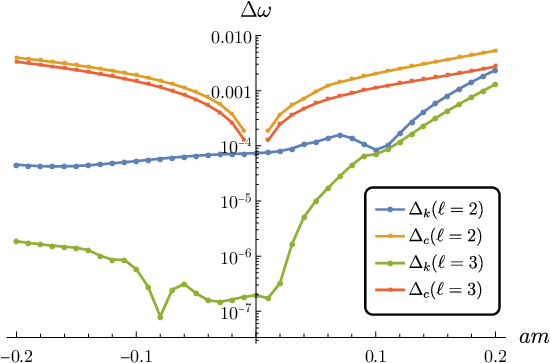}
\includegraphics[scale=0.80]{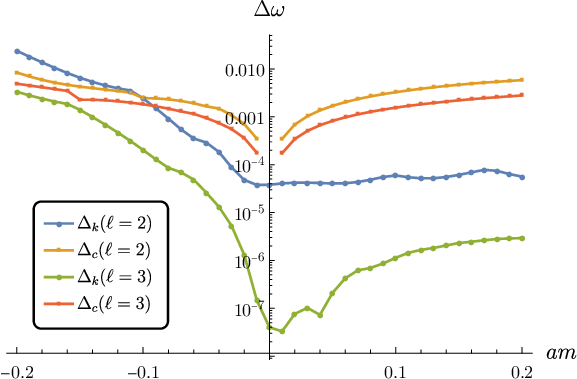}
	\caption{The comparison between the noncommutative correction $\Delta _c$ and the relative error $\Delta _k$ in the optimal WKB order is illustrated. Left: Axial case. Right: Polar case.}
\label{figerror}
\end{figure}

For commutative axial and polar potentials, the sixth-order WKB approximation usually yields the optimal result for the $\ell=2$ mode (see Table \ref{tab3} in appendix \ref{appc}). 
However, noncommutative QNM frequencies are optimized for different WKB orders (see Table \ref{tab4} -\ref{tab20}). We select the optimal frequencies for each $a m$ value (see Table \ref{tab1}-\ref{tab2}). In order to gauge the relative error inherent in the WKB method compared to the noncommutative (NC) correction, we introduce the absolute NC correction, denoted as $\Delta _c$,
\begin{equation}
    \Delta _{c} = | \omega _{NC} -\omega_C |,
\end{equation}
where again $\omega$ refers to fundamental mode frequency.
This absolute NC correction has been computed across a spectrum of $am$ values for both axial and polar cases and is compared with the relative error $\Delta _k$ inherent in the optimized WKB order, as shown in Fig \ref{figerror}. 
It is evident that for $\ell=3$ (and similarly for higher $\ell$ values), the error is negligible compared to the $\ell=2$ case. The error $\Delta _k$ becomes comparable with the NC correction $\Delta _c$ after a certain value of $am$ the positive/negative side of $a m$ axis for axial/polar modes is reached, with this claim being valid for both types of perturbation and all angular momentum channels analyzed. Within the domain of $a m$ where the WKB approximation proves effective, the impact of the  noncommutative parameter on QNM frequencies surpasses the error in the WKB approximation.
 However, for positive $am$ values where the WKB analysis breaks down, we can obtain the NC correction using other semi-analytic methods. 
We calculate the QNM frequencies using the P{\"o}schl-Teller and Rosen-Morse methods in the next section, thereby validating the corrections obtained from the WKB approach within the admissible domain of $am$.
%%%%%%%%%%%%%%%%%%%%%%%%%%%%%%%%%%%%%%%%
%%%%%%%%%%%%%%%%%%%%%%%%%%%%%%%%%%%%%%%
\begin{table}[h!]
\begin{tabular}{m{4em} m{5em} m {12em} m{3em} m{12em} m{8em}}
& $a m  $  & WKB  & Order & P\"oschl-Teller & Rosen-Morse  \\
\multirow{ 2}{*}{$\ell =2$ }&&&\\
\cmidrule{2-6}
& $-0.2 $ & 0.3775(61) - 0.0883(97) i & 6 & 0.382049 - 0.090320 i & 0.38335 - 0.08924 i \\
& $-0.1 $ & 0.3755(14) - 0.0887(70) i & 6 & 0.380114 - 0.090466 i & 0.38057 - 0.09008 i \\
& $-0.01$ & 0.3738(07) - 0.0888(92) i & 6 & 0.378454 - 0.090521 i & 0.37855 - 0.09044 i \\
& $-0.001$& 0.3736(38) - 0.0888(91) i & 6 & 0.378294 - 0.090520 i & 0.37838 - 0.09044 i \\
& $ 0   $ & 0.3736(19) - 0.0888(91) i & 6 & 0.378276 - 0.090520 i & 0.37837 - 0.09044 i \\
& $0.001$ & 0.3736(01) - 0.0888(91) i & 6 & 0.378258 - 0.090520 i & 0.37838 - 0.09042 i \\
& $ 0.01$ & 0.3734(33) - 0.0888(88) i & 6 & 0.378099 - 0.090518 i & 0.37836 - 0.09030 i \\
& $ 0.1 $ & 0.3715(87) - 0.0889(38) i & 4 & 0.376562 - 0.090455 i & 0.37756 - 0.08961 i \\
& $ 0.2 $ & 0.36(8345) - 0.08(8195) i & 4 & 0.375007 - 0.090238 i & 0.37611 - 0.08930 i \\
\multirow{ 2}{*}{$\ell =3$ }&&&\\
\cmidrule{2-6}
& $-0.2 $  & 0.602768 - 0.092381 i & 8  & 0.605558 - 0.093206 i & 0.60880 - 0.09161 i \\
& $-0.1 $  & 0.600920 - 0.092632 i & 10 & 0.603847 - 0.093344 i & 0.60489 - 0.09283 i \\
& $-0.01$  & 0.599573 - 0.092706 i & 8  & 0.602548 - 0.093361 i & 0.60271 - 0.09328 i \\
& $-0.001$ & 0.599456 - 0.092703 i & 8  & 0.602432 - 0.093358 i & 0.60257 - 0.09329 i \\
& $ 0   $  & 0.599443 - 0.092703 i & 8  & 0.602420 - 0.093358 i & 0.60257 - 0.09328 i \\
& $0.001$  & 0.599431 - 0.092702 i & 8  & 0.602407 - 0.093358 i & 0.60259 - 0.09327 i \\
& $ 0.01$  & 0.599318 - 0.092696 i & 12 & 0.602295 - 0.093355 i & 0.60272 - 0.09315 i \\
& $ 0.1 $  & 0.5983(44) - 0.0924(52) i & 4  & 0.601353 - 0.093234 i & 0.60302 - 0.09241 i \\
& $ 0.2 $  & 0.59(6979) - 0.09(1480) i & 4  & 0.600761 - 0.092919 i & 0.60246 - 0.09208 i \\
\multirow{ 2}{*}{$\ell =4$ }&&&\\
\cmidrule{2-6}
& $-0.2 $ & 0.812255 - 0.093889 i & 11 & 0.814338 - 0.094360 i & 0.81927 - 0.09259 i \\
& $-0.1 $ & 0.810458 - 0.094097 i & 12 & 0.812664 - 0.094503 i & 0.81421 - 0.09394 i \\
& $-0.01$ & 0.809279 - 0.094166 i & 12 & 0.811533 - 0.094537 i & 0.81175 - 0.09446 i \\
& $-0.001$& 0.809188 - 0.094164 i & 12 & 0.811442 - 0.094535 i & 0.81163 - 0.09447 i \\
& $ 0   $ & 0.809178 - 0.094164 i & 12 & 0.811433 - 0.094535 i & 0.81163 - 0.09446 i \\
& $0.001$ & 0.809169 - 0.094164 i & 12 & 0.811423 - 0.094535 i & 0.81167 - 0.09445 i \\
& $ 0.01$ & 0.809084 - 0.094159 i & 12 & 0.811338 - 0.094533 i & 0.81191 - 0.09433 i \\
& $ 0.1 $ & 0.8085(45) - 0.0939(66) i & 4  & 0.810770 - 0.094427 i & 0.81302 - 0.09361 i \\
& $ 0.2 $ & 0.808(397) - 0.093(169) i & 4  & 0.810860 - 0.094128 i & 0.81311 - 0.09332 i \\
\hline
\end{tabular}
	\caption{Table of NC axial QNMs for $n=0$, $M=1$ ($R=2$), and $\ell=2,3,4$, computed using the WKB higher-order method, P\"oschl-Teller (PT) method, and Rosen-Morse method. Order stands for the optimal WKB order. The estimated error of the WKB method is provided in parentheses. }\label{tab1}
\end{table}
%%%%%%%%%%%%%%%%%%%%%%%%%%%%%%%%%%%%%%%
%%%%%%%%%%%%%%%%%%%%%%%%%%%%%%%%%%%%%%
\begin{table}[h!]
\begin{tabular}{m{4em} m{5em} m {12em} m{3em} m{12em} m{8em}}
& $a m$  & WKB  & Order  & P\"oschl-Teller & Rosen-Morse  \\
\multirow{ 2}{*}{$\ell =2$ }&&&\\
\cmidrule{2-6}
& $-0.2$  & 0.3(80198) - 0.0(83646) i & 3 & 0.382642 - 0.097531 i & 0.38379 - 0.09648 i \\
& $-0.1$  & 0.37(4735) - 0.09(1148) i & 4 & 0.380292 - 0.093609 i & 0.38178 - 0.09230 i \\
& $-0.01$ & 0.3738(64) - 0.0892(07) i & 5 & 0.378475 - 0.090866 i & 0.37890 - 0.09050 i \\
& $-0.001$& 0.3736(58) - 0.0889(67) i & 5 & 0.378308 - 0.090622 i & 0.37845 - 0.09050 i \\
& $0$     & 0.3736(36) - 0.0889(40) i & 5 & 0.378290 - 0.090595 i & 0.37839 - 0.09051 i \\
& $0.001$ & 0.3736(13) - 0.0889(14) i & 5 & 0.378272 - 0.090567 i & 0.37836 - 0.09049 i \\
& $0.01$  & 0.3734(13) - 0.0886(75) i & 5 & 0.378109 - 0.090322 i & 0.37821 - 0.09023 i \\
& $0.1$   & 0.3718(88) - 0.0861(75) i & 6 & 0.376612 - 0.088102 i & 0.37741 - 0.08744 i \\
& $0.2$   & 0.3711(29) - 0.0836(58) i & 7 & 0.375215 - 0.085959 i & 0.37794 - 0.08379 i \\
\multirow{ 2}{*}{$\ell =3$ }&&&\\
\cmidrule{2-6}
& $-0.2$  & 0.60(4367) - 0.09(2871) i & 3  & 0.605859 - 0.096241 i & 0.60775 - 0.09528 i \\
& $-0.1$  & 0.600(941) - 0.093(756) i & 4  & 0.603790 - 0.094691 i & 0.60594 - 0.09361 i \\
& $-0.01$ & 0.599571 - 0.092828 i & 9  & 0.602530 - 0.093487 i & 0.60308 - 0.09321 i \\
& $-0.001$& 0.599456 - 0.092716 i & 10 & 0.602430 - 0.093377 i & 0.60263 - 0.09328 i \\
& $0$     & 0.599443 - 0.092703 i & 10 & 0.602419 - 0.093365 i & 0.60257 - 0.09329 i \\
& $0.001$ & 0.599431 - 0.092690 i & 10 & 0.602408 - 0.093353 i & 0.60255 - 0.09328 i \\
& $0.01$  & 0.599324 - 0.092577 i & 10 & 0.602314 - 0.093235 i & 0.60248 - 0.09315 i \\
& $0.1$   & 0.598589 - 0.091402 i & 9  & 0.601588 - 0.092150 i & 0.60286 - 0.09153 i \\
& $0.2$   & 0.598402 - 0.090106 i & 8  & 0.601184 - 0.091006 i & 0.60541 - 0.08897 i \\
\multirow{ 2}{*}{$\ell =4$ }&&&\\
\cmidrule{2-6}
& $-0.2$  & 0.813(422) - 0.094(494) i & 3  &  0.814702 - 0.096084 i & 0.81716 - 0.09519 i \\
& $-0.1$  & 0.8104(89) - 0.0947(78) i & 4  &  0.812601 - 0.095290 i & 0.81527 - 0.09432 i \\
& $-0.01$ & 0.809269 - 0.094238 i & 10 &  0.811514 - 0.094611 i & 0.81218 - 0.09437 i \\
& $-0.001$& 0.809187 - 0.094171 i & 12 &  0.811440 - 0.094548 i & 0.81170 - 0.09445 i \\
& $0$     & 0.809178 - 0.094164 i & 12 &  0.811432 - 0.094541 i & 0.81163 - 0.09447 i \\
& $0.001$ & 0.809170 - 0.094156 i & 12 &  0.811424 - 0.094534 i & 0.81161 - 0.09447 i \\
& $0.01$  & 0.809097 - 0.094088 i & 12 &  0.811357 - 0.094463 i & 0.81158 - 0.09438 i \\
& $0.1$   & 0.808733 - 0.093367 i & 11 &  0.810973 - 0.093790 i & 0.81270 - 0.09317 i \\
& $0.2$   & 0.808973 - 0.092531 i & 10 &  0.811068 - 0.093024 i & 0.81676 - 0.09100 i \\
\end{tabular}
\caption{Table of NC polar QNMs for $n=0$, $M=1$, and $\ell=2,3,4$, calculated using the WKB higher-order method, P\"oschl-Teller (PT) method, and Rosen-Morse method. Order stands for the optimal WKB order. The estimated error of the WKB method is provided in parentheses.} \label{tab2}
\end{table}

%%%%%%%%%%%%%%%%%%%%%%%%%%%%%%%%%%%%%%%%%%%%%%
%%%%%%%%%%%%%%%%%%%%%%%%%%%%%%%%%%%%%%%%%%%%%
\subsection{P{\"o}schl-Teller and Rosen-Morse method}
The method of deriving quasinormal modes from the knowledge of bound states of the inverted potential has been widely utilized in black hole perturbation studies. The P{\"o}schl-Teller potential was originally introduced as a solvable potential for which the Schr{\"o}dinger equation has exact solutions \cite{Poschl:1933zz}. This method has been adapted for black holes by approximating the more complex black hole potentials with the P{\"o}schl-Teller potential \cite{Mashhoon:1982im, Ferrari:1984ozr, Ferrari:1984zz, Blome:1981azp}. This approach involves approximating the effective potential in the Schr{\"o}dinger-like equation, derived from the radial perturbation equation, with the well-known P{\"o}schl-Teller potential. Consequently, the master wave equation can be rewritten as
\begin{equation} \label{PT_diff}
	\frac{\partial \Psi}{\partial r_*^2}+\left[ \omega ^2 -\frac{V_0}{\cosh ^2\alpha (r_*-  \bar r_*)} \right] \Psi =0,
\end{equation}
where
\begin{equation}
    \alpha ^2 =- \frac{V_0^{\prime \prime}}{2V_0}
\end{equation}
and $\bar r_*$ denotes the point where the effective potential reaches its maximum value in tortoise coordinate, and $V_0=V(\bar r_*)$ is the value of the potential at this point. A semi-analytical form for the quasinormal mode (QNM) frequencies can be obtained by transforming equation \eqref{PT_diff} into hypergeometric form and analyzing the asymptotic behavior of the solutions \cite{Berti:2009kk}. The formula for the frequencies of QNMs is given by
\begin{equation}\label{PTomega}
    \omega = \pm \sqrt{V_0-\frac{\alpha ^2}{4}}-i \alpha \left( n+\frac{1}{2} \right).
\end{equation}
From the above expression, it is clear that both the real and imaginary parts depend on $V_0$ and $V_0^{\prime \prime}$, similar to the WKB method. However, only the imaginary part of the quasinormal mode (QNM) is influenced by the overtone number $n$. Thus, this method is not recommended for determining the real part of QNM, except in specific cases such as the eikonal limit ($\ell \rightarrow \infty $) or for the fundamental mode $(n=0)$ \cite{Berti:2009kk}.

The effectiveness of this approach in approximating the effective potential can be enhanced by selecting a more accurate potential model to describe the black hole potential (see \cite{Boonserm:2010px} for an extensive list of potential options for QNM calculations). Here, we opt for a specific potential known as the Rosen-Morse potential \cite{Heidari:2023yjd, Heidari:2023egu}, which introduces a correction term to add asymmetry to the P{\"o}schl-Teller potential, resulting in a better correlation with the black hole effective potential. The Rosen-Morse potential function is defined as \cite{PhysRev.42.210}
\begin{equation}
	V_{RM}=   \frac{V_0}{\cosh ^2\alpha (r_* -  \bar r_*)} + V_1 \tanh \alpha (r_*-  \bar r_* ),
\end{equation}
where $V_1$ adds asymmetry to the potential.
By substituting this into the master wave equation and employing the appropriate boundary conditions, the resulting QNM frequencies are given by
\begin{equation}
 \frac{\sqrt{\omega ^2 +V_1}+\sqrt{\omega ^2 -V_1} }{ 2}  = \pm \sqrt{V_0-\frac{\alpha ^2}{4}}-i \alpha \left( n+\frac{1}{2} \right).
\end{equation}
When $V_1=0$ this reduces to \eqref{PTomega}. The obtained QNM frequencies using the P{\"o}schl-Teller and Rosen-Morse potentials are presented alongside the WKB method in Table \ref{tab1} and Table \ref{tab2}. 
These methods supplement our findings on noncommutative corrections calculated using the WKB method.

\section{Axial vs polar perturbations -- violation of classical isospectrality} \label{secV}

An important characteristic of quasinormal modes in classical general relativity is their isospectrality. For the commutative Schwarzschild black hole, axial and polar mode perturbations share identical spectra, despite being described by different master equations. This characteristic also extends to Reissner-Nordstr{\"o}m, Kerr, and Kerr-Newman black holes (to linear order in rotation) in the commutative case \cite{Pani:2013ija, Pani:2013wsa}. Chandrasekhar and Detweiler demonstrated for the commutative Schwarzschild case that the master equations of axial and polar perturbations are related by a specific transformation \cite{Chandrasekhar:1975zza}. This transformation was later identified as a subclass of the Darboux transformation \cite{Glampedakis:2017rar} (see Ref. \cite{darboux1882proposition, darboux1999proposition, Yurov:2018ynn} also). Isospectrality remains valid even in spacetimes including a cosmological constant for uncharged and non-rotating black holes \cite{Moulin:2019bfh}. However, the physical origin of this property remains unclear, and it is uncertain whether isospectrality is a generic feature or a mere coincidence for classical black holes. Efforts to understand isospectrality in black hole perturbation remain intriguing, particularly since this property holds only for very specific black hole spacetimes.  \\

Notably, in classical general relativity, there are cases where isospectrality is broken in black hole perturbations \cite{Ferrari:2000ep, Cardoso:2001bb, Berti:2003ud, Blazquez-Salcedo:2019nwd}. Isospectral breaking is also observed in beyond general relativity theories \cite{Bhattacharyya:2017tyc, Bhattacharyya:2018qbe, Datta:2019npq, Bhattacharyya:2018hsj, Kobayashi:2012kh, Kobayashi:2014wsa, del-Corral:2022kbk, Chen:2021pxd}.
%such as f(R) gravity \cite{Bhattacharyya:2017tyc, Bhattacharyya:2018qbe, Datta:2019npq}, dynamical Chern-Simons gravity \cite{Bhattacharyya:2018hsj}, scalar-tensor theories \cite{Kobayashi:2012kh, Kobayashi:2014wsa}, black holes in higher-dimensional scenarios \cite{Berti:2009kk}, cases of loop quantum gravity \cite{del-Corral:2022kbk} and in nonlocal model of modified gravity \cite{Chen:2021pxd} etc. 
In fact, parameterization approaches have shown that the isospectral property between even- and odd-parity quasinormal mode spectra is quite fragile \cite{Cardoso:2019mqo}. Therefore, any observational indication of isospectrality breaking would strongly suggest new physics beyond general relativity \cite{Shankaranarayanan:2019yjx}. From a quantum gravity perspective, it has even been suggested that isospectral breaking could potentially signal quantum gravity observables in current experiments \cite{Liu:2024oeq}. \\

Classically, the isospectrality can be understood as follows. The master equations are of the form
\begin{equation*}
	\frac{d^2 \psi}{d r_*} + \big(\omega^2 - V_\pm(r)\big)\psi = 0,
\end{equation*}
where $+$ and $-$ correspond to the polar (even) and axial (odd) perturbations respectively, governed by the commutative Zerilli and Regge-Wheeler potentials,
\begin{align*}
	V_+(r) = \ & V_Z(r) =  \frac{(r-R) \left(8 \Lambda ^2 (\Lambda +1) r^3+12 \Lambda ^2 r^2 R+18 \Lambda  r R^2+9 R^3\right)}{r^4 (2 \Lambda  r+3 R)^2},\\
	V_-(r) = \ & V_{RW}(r) =  \frac{(r-R)(2r(1 + \Lambda) - 3R)}{r^4},
\end{align*}
with the unique tortoise coordinate given by
\begin{align*}
	r_* =& \ r + R \log \frac{r - R}{R}.
\end{align*} 
The criterion to test whether two potentials are equivalent via the Darboux transformation is given by \cite{Glampedakis:2017rar}
\begin{equation*}
	\frac{\frac{d}{d r_*}(V_+ + V_-)}{V_+ - V_-} = \int dr_* (V_+ - V_-),
\end{equation*}
where the potentials should be expressed using the standard tortoise coordinate $r_* = r + R \log \frac{r-R}{r}$.
This condition is satisfied for the $V_+$ and $V_-$ in the commutative case. 
One of the requirements of the Darboux transformation is that initially, neither the polar nor the axial equations contain first derivative terms and that they have the same tortoise coordinate.
In the noncommutative case, tortoise coordinates $\hat r_*$ differ for axial and polar modes as can be seen from equations \eqref{RWtortoise} and \eqref{Ztortoise}.
Trying to express one tortoise coordinate in terms of the other in any of the two equations \eqref{Rschrodinger} and \eqref{Zschrodinger} spoils their Schr\"odinger form by introducing a first derivative term in one of the equations.
It is therefore impossible to relate them through the Darboux transformation since initial requirements cannot be satisfied, hinting at breaking of the classical isospectrality.
The isospectrality violation is evident from the numerical values of QNM frequencies. In the commutative case $(am=0)$, the QNM frequencies are identical for both the axial and polar cases (Table \ref{tab0a}).\\
\begin{table}[h!]
\begin{tabular}{m{4em} m{12em} m{12em} m {12em} }
%\hline
\\
 & $\ell=2$ & $\ell=3$ &  $\ell=4$ \\
%\\
\hline
\\
Axial & 0.3736(19) - 0.0888(91) i & 0.599443 - 0.092703 i & 0.809178 - 0.094164 i \\
Polar & 0.3736(36) - 0.0889(40) i & 0.599443 - 0.092703 i & 0.809178 - 0.094164 i \\
\\
\hline
\end{tabular}
\caption{QNM frequencies calculated using the WKB method for commutative case $(am=0)$ with  $n=0$, $M=1\ (R=2)$.} 
\label{tab0a}
\end{table}
\par
As we introduce noncommutativity with a small value, such as $am=-0.001$, the breaking of isospectrality becomes apparent in higher $\ell$ modes (\ref{tab0b}).
\\
\begin{table}[h!]
\begin{tabular}{m{4em} m{12em} m{12em} m {12em} }
%\hline
\\
 & $\ell=2$ & $\ell=3$ &  $\ell=4$ \\
%\\
\hline
\\
Axial & 0.3736(38) - 0.0888(91) i & 0.599456 - 0.092703 i & 0.809188 - 0.094164 i \\
Polar & 0.3736(58) - 0.0889(67) i & 0.599456 - 0.092716 i & 0.809187 - 0.094171 i \\
\\
\hline
\end{tabular}
\caption{QNM frequencies calculated using the WKB method for the noncommutative case $am=-0.001$ with  $n=0$, $M=1\ (R=2)$.} 
\label{tab0b}
\end{table}
\par
As we further increase the strength of noncommutativity, for instance to $am=-0.01$, the breaking of isospectrality becomes more pronounced, being manifest even in the fundamental mode $\ell=2$ (see Table \ref{tab0c}).\\
\begin{table}[h!]
\begin{tabular}{m{4em} m{12em} m{12em} m {12em} }
%\hline
\\
 & $\ell=2$ & $\ell=3$ &  $\ell=4$ \\
%\\
\hline
\\
Axial & 0.3738(07) - 0.0888(92) i & 0.599573 - 0.092706 i & 0.809279 - 0.094166 i \\
Polar & 0.3738(64) - 0.0892(07) i & 0.599571 - 0.092828 i & 0.809269 - 0.094238 i \\
\\
\hline
\end{tabular}
\caption{QNM frequencies calculated using the WKB method for the noncommutative case $am=-0.01$ with $n=0$, $M=1\ (R=2).$} 
\label{tab0c}
\end{table}
\par
\begin{figure}[h!]
\centering
\subfigure[ref2][]{\includegraphics[scale=0.8]{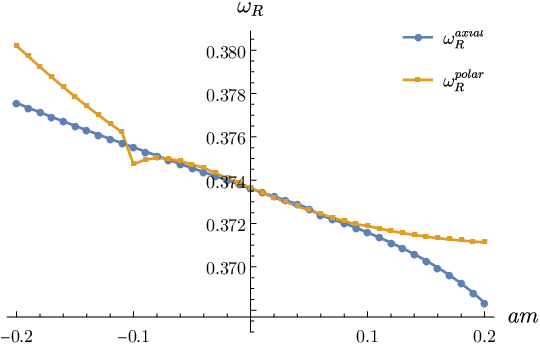}}
\qquad
\subfigure[ref3][]{\includegraphics[scale=0.8]{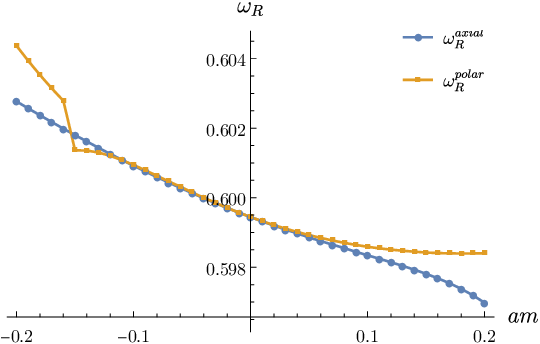}}
  \subfigure[ref3][]{\includegraphics[scale=0.8]{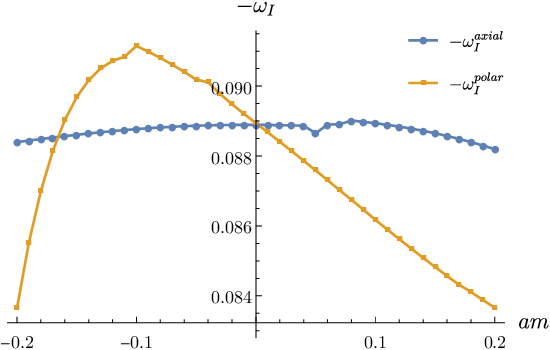}}
\qquad
\subfigure[ref3][]{\includegraphics[scale=0.8]{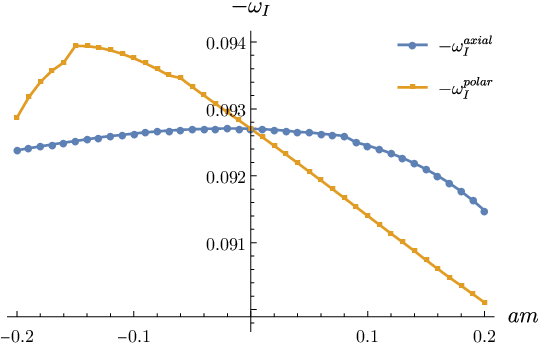}}
\subfigure[ref2][]{\includegraphics[scale=0.8]{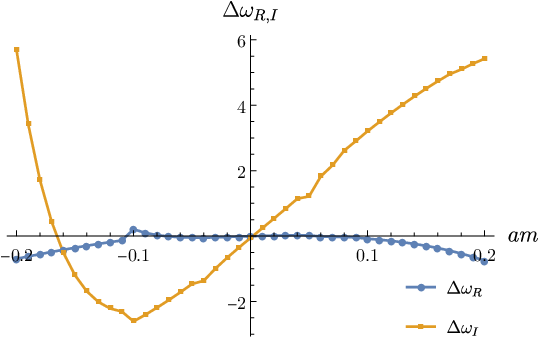}}
\qquad
\subfigure[ref3][]{\includegraphics[scale=0.8]{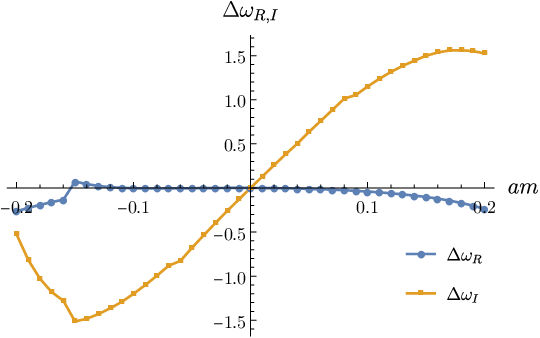}}
\caption{The figures depict the breaking of the isospectrality caused by noncommutativity. The left panel corresponds to $\ell=2$, while the right panel corresponds to $\ell=3$. In the top row, the variation of the real part of the QNM $(\omega _R)$ with $am$ is shown. In the middle row, the variation of the imaginary part of the QNM $(\omega _I)$ with $am$ is depicted. In the bottom row, the relative deviation between the two modes, $\Delta \omega _{R,I}$, is illustrated. The parameters fixed in these plots are $n=0$, and $M=1$ ($R=2$). In each case, the QNM values corresponding to the optimal WKB order are chosen.}
\label{figiso6}
\end{figure}

The increasing effect of isospectral breaking continues to appear for all negative values of $am$, and it also becomes evident on the positive side of $am$. We assess the degree of isospectral breaking as follows \cite{Liu:2024oeq}, 

\begin{equation}
    \Delta \omega _{R,I}=100 \times \frac{\omega _{R,I}^{axial}-\omega _{R,I}^{polar}}{\omega _{R,I}^{polar}}.
\end{equation}
The results are shown in Figure \ref{figiso6}. 
Firstly, it is visible that the real part of the QNM frequency is affected similarly in both axial and polar cases, with deviations appearing at larger parameter values where the precision of the WKB method starts to decrease.
Secondly, the negative imaginary part of QNM, $\omega _I$, is influenced by $am$ in the reverse order compared to the real part. However, it should be noted that in the axial case, $\omega _I$ is less influenced by noncommutativity. These observations imply that the effects of noncommutativity on oscillation frequencies and damping times differ in magnitude for axial and polar modes. The $\Delta \omega _{R,I}$ depicted in the figure shows that the imaginary part experiences a more pronounced isospectral breaking effect than the real part.\\

The fact that polar modes are significantly influenced by the noncommutativity stemming from the twist defined in \eqref{kxdef} can be qualitatively analyzed as follows. Axial perturbations, which parameterize the off-diagonal components of the perturbation matrix, describe the twisting of spacetime around the black hole. These typically represent the gravitational waves that could be generated by the rotating or spiraling motion of matter in the vicinity of the black hole. \\

In contrast, polar perturbations impact the scalar and spherical components of the metric, modifying distances while preserving the overall spherical symmetry. These perturbations include changes in the radial and temporal components of the metric, representing the expansion and contraction of spacetime, or the 'breathing-like' modes of the black hole. \\

Given that polar perturbations predominantly relate to the radial direction, and considering that it is the radial coordinate that is quantized by the field $X = \partial_r$ in the twist \eqref{kxdef}, it is understandable why polar perturbations are more sensitive to the coordinate quantization induced by the twist.\\

  Polar modes exhibit both higher and lower imaginary QNM frequencies compared to axial modes, as seen in Figure \ref{figiso6}. Specifically, for negative values of $am$, the polar modes have larger imaginary frequencies (implying stronger damping), while for positive values of $am$, they have smaller imaginary frequencies (weaker damping). This could lead to a more striking statistical data output on the polar modes side recorded at the gravitational wave detector, provided that the detector can distinguish between the polar and axial modes.
Similarly, if the detector is able to distinguish azimuthal numbers of the waves and measure that $m$ and $-m$ have different damping times, this could hint at $r - \varphi$ quantization. \\

Expressions for the tortoise coordinate and effective potential become similar for noncommutative axial and polar modes when $\ell \to \infty$:
\begin{equation} \label{nceikonal}
	\hat r_* = r + R \log \frac{r - R}{R} \pm \frac{\lambda a}{2} \frac{R}{r - R}, \qquad 
	\hat{V}(r) = \frac{\ell^2 (r - R)}{r^3} \pm \frac{\lambda a}{2} \frac{\ell^2 (3R - 2r)}{r^4},
\end{equation}

where $+$ corresponds to axial and $-$ to polar perturbations.
A natural question to ask is whether the spectrum corresponding to these potentials differs from the commutative one. Consider the limit $a \to 0$ of the expressions above:
\begin{equation} \label{eikonal}
	r_* = r + R \log \frac{r - R}{R}, \qquad
	V(r) = \frac{\ell^2 (r - R)}{r^3}. 
\end{equation}
Replacing $r \to r \pm \lambda a / 2$ here and expanding up to first order in $a$ turns the RHS parts of these relations into RHS parts in the equations \eqref{nceikonal}.
Consequentially, solutions of \eqref{nceikonal} can be obtained as translated solutions of \eqref{eikonal}, provided boundary conditions are translated accordingly.
Therefore the noncommutative QNM spectrum, up to the first order in $a$, becomes identical to the commutative one in the eikonal limit. \\

The physical significance of breaking isospectrality is profound. In the context of gravitational wave observations, differing QNM frequencies can serve as a tool to distinguish between different gravitational theories. For instance, if future gravitational wave detections from black hole mergers reveal discrepancies in the expected QNM frequencies between axial and polar modes, it could indicate the presence of new physics beyond general relativity. Such observations may provide empirical tests of modified theories of gravity and offer insights into the fundamental nature of spacetime.
Consequently, experimental bounds on isospectrality are directly tied to a possibility of deviating from general relativity. Detecting any such deviations in gravitational wave observations would provide critical information about the underlying quantum gravity effects, thus paving the way for new theoretical developments and a deeper understanding of the universe's structure.

\section{Outlook and discussion}

A generic black hole  can be described completely by a small number of intrinsic parameters such as the mass, charge and spin of a black hole and in the case of somewhat more general theoretical framework such as that of beyond Einsteinian gravity, this description may include some additional parameters  characterizing  the theory. \\

When disturbed, a black hole tends to go back to its equilibrium characterized by a nonperturbed black hole solution. In turn, a response of a black hole to external perturbation is characterized by a discrete set of quasinormal mode frequencies which can also be described by a small set of intrinsic  parameters of the theory. This turns QNMs into preferable  instrument for accomplishing at least two important tasks: to reveal the information about parameters of the black hole in the ringdown phase resulting from a binary black hole merger \cite{Dreyer:2003bv} and to test theories of gravity beyond Einstein, as these  vibration  frequencies exhibit a universal part of the gravitational wave signals and above all can be measured \cite{Dreyer:2003bv, Franchini:2023eda}. \\

In this paper, we have examined a specific type of Einstein gravity modification based on the framework of noncommutative geometry and quantization methods implemented by the Drinfeld twist. This approach takes into account that  the geometry of spacetime acquires a granular structure at the Planck scale distances and is  particularly well suited for implementing this feature into a formalism. \\

The analysis has been made in a bottom-up approach, implying that the equation of motion is obtained by carefully revising the basic notions of differential geometry, such as the vector field, Lie derivative, covariant derivative, etc, rather than from some anticipated overruling/paramount action principle. In particular, a deformation quantization by a Drinfeld twist operator has been applied  to a linearized gravitational perturbation theory. After developing  a general perturbation theory up to linear order in both the deformation parameter and metric perturbation, we  have applied it to a fixed Schwarzschild background in order to analyze the corresponding axial as well as polar type of gravitational perturbations. In this way, we have derived the noncommutative generalizations of the Regge-Wheeler and Zerilli potentials, respectively. 
  It should be noted that equations leading up to effective potential are valid for two sources of noncommutativity (and their linear combinations): $r - \varphi$ and $r - t$. The analysis of the spectrum has been carried out only for the $r - \varphi$ case. \\

The Abelian Drinfeld twist results in noncommutativity of Moyal type, which features simplicity, generality, and versatility at low-energy regimes, where it reflects general features of NC spacetime. Specifically, the semi-Killing form of the twist introduced non-trivial NC corrections in the perturbations around Schwarzschild spacetime while ensuring separability of equations. Alternative coordinate choices, involving non-constant functions, can still be expressed in a Moyal form when suitably adapted, thus emphasizing the versatility of Moyal twist-based approaches. 
Ultimately, our selected twist demonstrates explicitly how noncommutativity could modify gravitational wave signals, providing a specific, testable prediction and an important starting point for future experimental and theoretical inquiries into NC gravitational physics. \\

For both potentials QNM spectrum has been calculated to a high precision in WKB, which has enabled us to investigate possible isospectrality breaking. Namely, isospectrality between even- and odd-parity perturbations of a Schwarzschild black hole and some more general geometries  is a well-known property of classical general relativity.
We have shown  that this property becomes violated when the quantum structure of spacetime is assumed. This is consistent with other findings in the literature that were inferred for some effective models originating in loop quantum gravity \cite{del-Corral:2022kbk,Cruz:2020emz,Cruz:2015bcj}. Interestingly however, isospectrality appears to be restored in the eikonal limit of large angular momentum.\\

  A deeper insight into the spectrum and nature of the noncommutative metric perturbations can be gained by employing more robust methods, such as Leaver's continued fraction method or direct integration, and by considering the time profile of the perturbations. It has been recently shown that even small perturbations to the spectrum can lead to instability \cite{Courty:2023rxk}. We plan to address this in a future work. Preliminary steps have already been taken in \cite{Herceg:2024upt}, where we showed that another radial coordinate can be used to make the behaviour of the tortoise coordinate and effective potential more similar to commutative case in the near-horizon region, which is a necessary starting point for these more advanced methods.\\

The way how a black hole responds to external perturbation is of essential importance, as these response patterns  influence  the  gravitational wave emission. As a consequence, they are generally measurable. In this paper, we have studied the linear response of a black hole to small perturbations, as described  by  the quasinormal mode frequencies, which describe how a black hole relaxes back to the Schwarzschild solution upon being disturbed. \\

Another type of linear response that would be of interest to study is the linear response of a black hole to a long-wavelength tidal gravitational field. This type of response is expressed in terms of tidal Love numbers \cite{Love}, which outline the feedback of a rigid body to an external deformation and thus represent a measure of the rigidity of the object being studied. In principle, the feedback of the black hole to an external field could  also be studied in  case the deformation is caused by the background electromagnetic and scalar field profiles. \\

As the gravitational wave signal coming from an orbiting binary system depends on the tidal Love numbers\footnote{It appears that the gravitational wave signal is affected by the tidal Love numbers at subleading post-Newtonian order.}  of the constituents, the latter may be inferred by measuring gravitational waves during the inspiral phase \cite{Flanagan:2007ix}. In our case it would be interesting to see how quantum spacetime deformation affects the tidal Love numbers of the Schwarschild black hole \cite{Hui:2020xxx}. Another point of interest would be to extend the results for the QNM spectra obtained here to Kerr black holes, which we plan to address in an upcoming work.  \\

\noindent{\bf Acknowledgment}\\
This  research was supported by the Croatian Science Foundation Project No. IP-2020-02-9614 \textit{Search for Quantum spacetime in Black Hole QNM spectrum and Gamma Ray Bursts}. Part of the calculations were checked using the Mathematica package.

\newpage
\appendix

\section{Zerilli gauge } \label{Zerilli_gauge_app}
%\frametitle{}
The polar perturbations of a static spherically symmetric black hole are parameterized by seven families of functions: $H_{0}^{\ell m}$, $H_{1}^{\ell m}$, $H_{2}^{\ell m}$, $\alpha^{\ell m}$, $\beta^{\ell m}$, $K^{\ell m}$, and $G^{\ell m}$, dependent on the variables $(r,t)$ \cite{Regge:1957td, Langlois:2021xzq}. The non-zero components of the metric perturbation $h_{\mu \nu}$ are expressed as follows:
\begin{eqnarray}
\label{eq:even-pert1}
&&h_{tt} = A(r) \sum_{\ell, m} H_{0}^{\ell m}(t, r)  Y_{\ell m}(\theta,\varphi), \quad
h_{tr} = \sum_{\ell, m} H_{1}^{\ell m}(t,r) Y_{\ell m}(\theta,\varphi),\\
&&h_{rr} = \frac{1}{A(r)} \sum_{\ell, m} H_{2}^{\ell m}(t,r) Y_{\ell m}(\theta,\varphi), \\
&& h_{ta} = \sum_{\ell, m} \beta^{\ell m}(t,r) \partial_a Y_{\ell m}(\theta,\varphi), \quad
h_{ra} = \sum_{\ell, m} \alpha^{\ell m}(t,r) \partial_a Y_{\ell m}(\theta,\varphi), \\
\label{eq:even-pert4}
&&h_{ab} = \sum_{\ell, m} K^{\ell m}(t,r) g_{ab} Y_{\ell m}(\theta,\varphi) + \sum_{\ell, m} G^{\ell m}(t,r) D_a D_b Y_{\ell m}(\theta,\varphi)\, ,
\end{eqnarray}
where $A(r)=1-R/r$. The indices $a$ and $b$ take values of angular coordinates $\theta$ and $\varphi$. Here, $D_a$ is the 2-dimensional covariant derivative associated with the metric of the 2-sphere $d\theta ^2 + \sin^2 \theta d \varphi ^2$. The explicit form of the angular part of the metric $h_{\mu \nu}$ is given by
	\begin{align}
		h_{\theta\theta} &=  \sum_{\ell, m} K^{\ell m}(t,r) Y_{\ell m}(\theta,\varphi) + \sum_{\ell, m} G^{\ell m}(t,r) \partial_{\theta}^2Y_{\ell m}(\theta,\varphi)  \,, \\
		h_{\theta\varphi} &=h_{\varphi\theta} = - \sum_{\ell, m} G^{\ell m}(t,r) \cotan\theta \, \partial_\varphi Y_{\ell m}(\theta,\varphi) \,, \\
		h_{\varphi\varphi} &= \sum_{\ell, m} \sin^2\theta \, K^{\ell m}(t,r) Y_{\ell m}(\theta,\varphi) + \sum_{\ell, m} G^{\ell m}(t,r)\left(\partial_{\varphi}^2 + \sin\theta\cos\theta \, \partial_\theta\right)Y_{\ell m}(\theta,\varphi) \,.
	\end{align}
Due to the spacetime diffeomorphism invariance of the theory, the parametrization of perturbations presented above is redundant and can be simplified through gauge fixing. At the linear level, an infinitesimal change of coordinates, $x^\mu \rightarrow x^\mu + \xi^\mu$, induces the transformation:
\begin{eqnarray}
   \label{diffeo}
h_{\mu\nu} \rightarrow h_{\mu\nu} + \nabla_\mu \xi_\nu + \nabla_\nu \xi_\mu .
\end{eqnarray}
 In the polar sector, one can choose the gauge parameter $\xi$ in such a way that three families of functions in (\ref{eq:even-pert1} - \ref{eq:even-pert4}), namely $G^{\ell m}$, $\alpha^{\ell m}$, and $\beta^{\ell m}$, are set to zero. This particular gauge is known as the Zerilli gauge. In the Zerilli gauge, the polar perturbations can be parametrized by the remaining four families of functions, namely $H_{0}^{\ell m}$, $H_{1}^{\ell m}$, $H_{2}^{\ell m}$, and $K^{\ell m}$, as follows:
\begin{eqnarray}
&&h_{tt} = A(r)\sum_{\ell, m} H_{0}^{\ell m}(t,r) Y_{\ell m}(\theta,\varphi), \quad
h_{tr} = \sum_{\ell, m} H_{1}^{\ell m}(t,r) Y_{\ell m}(\theta,\varphi),\\
&&h_{rr} = \frac{1}{A(r)} \sum_{\ell, m} H_{2}^{\ell m}(t,r) Y_{\ell m}(\theta,\varphi), \quad 
h_{ab} = \sum_{\ell, m} K^{\ell m}(t,r) g_{ab} Y_{\ell m}(\theta,\varphi) .
\end{eqnarray}

\section{Coefficients of transformation} \label{app-coeff}
In this appendix we present the coefficients $\{ f(r), g(r), h(r), l(r)\}$ and $\{\tilde f(r), \tilde g(r), \tilde h(r), \tilde l(r)\}$ that constitute the transformation equation \eqref{transformation}. Together with the equation \eqref{Ztortoise}, the field redefinition satisfies the constraints outlined in \eqref{trans-constraints}. 
The functions appearing in Eq. \eqref{eqfghl} are given by
\begin{equation}
    \begin{aligned}
	g(r) = &1,\\
        f(r)= &\frac{2 \Lambda  (\Lambda +1) r^2+3 \Lambda  r R+3 R^2}{r^2 (2 \Lambda  r+3 R)}, \\
        l(r) = & \frac{i r^2}{R-r},\\
        h(r) = & \frac{3 i R}{2 \Lambda  r+3 R}-\frac{i (2 r-3 R)}{2 (r-R)},
    \end{aligned}
\end{equation}
which are the commutative results derived in Ref. \cite{Zerilli:1970se} and the non commutative corrections are
\begin{equation}
    \begin{aligned}        
        \tilde f(r) = & \, \frac{1}{72 r^3 (2 \Lambda  r+3 R)^2} \Biggl\{\frac{36}{r-R} \biggl(-4 \Lambda  (\Lambda  (3 \Lambda +5)+5) r^4+8 (\Lambda -1) \Lambda  (2 \Lambda +1) r^3 R \Biggr. \Biggr. \\ 
        & \Biggl.  \Biggl. +(\Lambda  (44 \Lambda +21)-36) r^2 R^2+(32 \Lambda +57) r R^3-6 R^4\biggr)- \biggl( r (2 \Lambda  r+3 R) \left(4 \Lambda  (\Lambda +1) r^2+6 \Lambda  r R+6 R^2\right) \biggr) \Biggr. \\
        & \Biggr. \times \left[\frac{9 \left(8 \Lambda ^2+34 \Lambda +9\right) \log (r-R)+4 \Lambda  (\Lambda +3) (10 \Lambda -9) \log (2 \Lambda  r+3 R)}{(2 \Lambda +3)^2 R} \right. \Biggr. \\
        & \Biggl. \left. +\frac{36}{2 \Lambda +3} \left(\frac{1}{r-R}-\frac{3}{2 \Lambda  r+3 R}\right)-\frac{(10 \Lambda +9) \log (r)}{R}-\frac{123}{r}\right]\Biggr\}, \\
        \tilde l(r) = & -\frac{i  r }{36 (r-R)} \Biggl\{ \frac{-9 \left(8 \Lambda ^2+34 \Lambda +9\right) r \log (r-R)-4 \Lambda  (\Lambda +3) (10 \Lambda -9) r \log (2 \Lambda  r+3 R)}{(2 \Lambda +3)^2 R} \Biggr. \\
        &\Biggl. -\frac{36 r}{2 \Lambda +3} \left(\frac{1}{r-R}-\frac{3}{2 \Lambda  r+3 R}\right)+\frac{(10 \Lambda +9) r \log (r)}{R}+69\Biggr\},\\
        \tilde h(r) = & \frac{i  }{72 (2 \Lambda +3)^2 r R (r-R)^2 (2 \Lambda  r+3 R)^2}\Biggl\{ \biggl( 2 r (r-R) (2 \Lambda  r+3 R) \left(3 R (2 \Lambda  r+R)-4 \Lambda  r^2\right) \biggr) \Biggr. \\
        & \Biggl. \times \left[\Lambda  \left(\left(20 \Lambda ^2+78 \Lambda +99\right) \log (r)-9 (4 \Lambda +17) \log (r-R)-2 (\Lambda +3) (10 \Lambda -9) \log (2 \Lambda  r+3 R)\right) \Biggr. \right.\\
        & \Biggl. \left. +81 \tanh ^{-1}\left(\frac{R}{2 r-R}\right)\right]+3 (2 \Lambda +3) R \Biggl[ 8 \Lambda  (\Lambda  (74 \Lambda +183)+72) r^4-4 \Lambda  (\Lambda  (202 \Lambda +69)-531) r^3 R \Biggr. \Biggr. \\
        & \Biggl. \Biggl. +12 (\Lambda  (\Lambda  (22 \Lambda -189)-213)+252) r^2 R^2+9 (2 \Lambda  (64 \Lambda -89)-579) r R^3+675 (2 \Lambda +3) R^4\Biggr]\Biggr\},\\
        \tilde g(r) = & -\frac{1}{36}  \Biggr\{\frac{1}{(2 \Lambda +3)^2 R}\left[ 9 \left(8 \Lambda ^2+34 \Lambda +9\right) \log (r-R)+4 \Lambda  (\Lambda +3) (10 \Lambda -9) \log (2 \Lambda  r+3 R) \right]\Biggl. \\
        & \Biggr.  +\frac{36}{2 \Lambda +3} \left(\frac{1}{r-R}-\frac{3}{2 \Lambda  r+3 R}\right)-\frac{(10 \Lambda +9) \log (r)}{R}-\frac{123}{r}\Biggl\} .
    \end{aligned}
\end{equation}

\newpage
\section{Error estimation in WKB approximation method} \label{appc}
In this appendix, we provide the details of error estimation in the QNM calculation using the higher-order WKB method for both axial and polar modes. Error for the $n$-th order WKB is equal to $| \omega_{n+1}-\omega_{n-1}|/2$ \cite{Konoplya:2019hlu, Konoplya:2004ip}. For the commutative case $(am=0)$ with $\ell=2$ we obtain
%%%%%%%%%%%%%%%%%%%%%%%%%%%%%%%%%%%%%%%%%%%%%%
%%%%%%%%%%%%%%%%%%%%%%%%%%%%%%%%%%%%%%%%%%%%%%%%
\begin{table}[h!]
\begin{tabular}{m{7em} m{12em} m {8em} m{12em} m{8em}}
\hline
\\
WKB  order & Axial  QNM  & Error &  Polar  QNM   &  Error\\
\\
\hline
\\
12 & 0.373913 - 0.090034 i & 0.001916 & 0.372655 - 0.089826 i & 0.001727 \\
11 & 0.374467 - 0.089901 i & 0.00061 & 0.374128 - 0.089473 i & 0.000855 \\
10 & 0.374217 - 0.088853 i & 0.000564 & 0.373945 - 0.088705 i & 0.000451 \\
9 & 0.373894 - 0.088930 i & 0.000191 & 0.373519 - 0.088806 i & 0.000254 \\
8 & 0.373850 - 0.088746 i & 0.00016 & 0.373578 - 0.089057 i & 0.000151 \\
7 & 0.373599 - 0.088806 i & 0.000136 & 0.373730 - 0.089020 i & 0.000093 \\
6 & 0.373619 - 0.088891 i & {\bf 0.000074} & 0.373707 - 0.088923 i & 0.000062 \\
5 & 0.373504 - 0.088918 i & 0.000121 &  0.373636 - 0.088940 i & {\bf 0.000038} \\
4 & 0.373553 - 0.089124 i & 0.000227 &0.373640 - 0.088959 i & 0.000323 \\
3 & 0.373162 - 0.089217 i & 0.002502 & 0.373012 - 0.089109 i & 0.002622 \\
\\
\hline
\end{tabular}
\caption{Error estimation in WKB method for commutative case $(am=0)$ with  $n=0$, $M=1 (R=2)$ and $\ell=2$.} 
\label{tab3}
\end{table}

From the table we conclude that the optimal WKB order for axial case is 6 whereas for the polar case, it is 5. The relative error between the WKB orders is of the order $10^{-4}$ in both cases.  \\ \\ \\ \\ \\
%%%%%%%%%%%%%%%%%%%%%%%%%%%%%%%%%%%%%%%%%%%%%%
%%%%%%%%%%%%%%%%%%%%%%%%%%%%%%%%%%%%%%%%%%%%%%%%
For the noncommutative case $am=-0.2$ with $\ell=2$ we obtain
\begin{table}[h!]
\begin{tabular}{m{7em} m{12em} m {8em} m{12em} m{8em}}
\hline
\\
WKB  order & Axial  QNM  & Error &  Polar  QNM   &  Error\\
 \\
\hline
\\
 12 & 0.375584 - 0.089656 i & 0.001044 &  \\
 11 & 0.377608 - 0.089176 i & 0.001082 &  \\
 10 & 0.377472 - 0.088598 i & 0.000318 &  \\
 9 & 0.377693 - 0.088546 i & 0.000157 &  \\
 8 & 0.377644 - 0.088335 i & 0.000118 & \\
 7 & 0.377552 - 0.088356 i & 0.000052 & \\
 6 & 0.377561 - 0.088397 i &{\bf 0.000046} & \\
 5 & 0.377482 - 0.088416 i & 0.000118 & 0.432880 - 0.288129 i & 0.437075 \\
 4 & 0.377533 - 0.088632 i & 0.000242 & 0.336591 - 0.094483 i & 0.10558 \\
 3 & 0.377115 - 0.088730 i & 0.002582 & 0.380198 - 0.083646 i & {\bf 0.023975} \\
 \\
\hline
\end{tabular}
\caption{Error estimation in WKB method for the noncommutative case $am=-0.2$ with  $n=0$, $M=1\ (R=2)$ and $\ell=2$. Orders for which the error exceeded the obtained value were omitted.} \label{tab4}
\end{table}

From the table we conclude that the optimal WKB order for the axial case is 6, whereas for the polar case it is 3. The relative error between the WKB orders is of the order of $10^{-4}$ in axial and $10^{-1}$ for the polar case. We observed that the higher-order WKB method breaks down after reaching certain orders in the polar case. This phenomenon appears to be a generic feature, particularly noticeable for higher (negative) values of the noncommutative parameter $am$ in the polar case.

\pagebreak
%%%%%%%%%%%%%%%%%%%%%%%%%%%%%%%%%%%%%%%%%%%%%%
%%%%%%%%%%%%%%%%%%%%%%%%%%%%%%%%%%%%%%%%%%%%%%%%
For the noncommutative case $am=-0.1$ with $\ell=2$ we obtain
\begin{table}[h!]
\begin{tabular}{m{7em} m{12em} m {8em} m{12em} m{8em}}
\hline
\\
WKB  order & Axial  QNM  & Error &  Polar  QNM   &  Error\\
 \\
\hline
\\
 12 & 0.375160 - 0.090027 i & 0.00085  &  \\
 11 & 0.376112 - 0.089799 i & 0.000666 &  \\
 10 & 0.375903 - 0.088921 i & 0.000457 & \\
 9  & 0.375766 - 0.088954 i & 0.000149 & \\
 8  & 0.375705 - 0.088699 i & 0.000165 & \\
 7  & 0.375509 - 0.088745 i & 0.000102 & 0.380796 - 0.032964 i & 0.198099 \\
 6  & 0.375514 - 0.088770 i & {\bf 0.000053} & 0.390405 - 0.092178 i & 0.031484 \\
 5  & 0.375414 - 0.088793 i & 0.000105 & 0.375878 - 0.095740 i & 0.007852 \\
 4  & 0.375456 - 0.088970 i & 0.00022  & 0.374735 - 0.091148 i & {\bf 0.002431} \\
 3  & 0.375066 - 0.089063 i & 0.002525 & 0.375851 - 0.090878 i & 0.003284 \\
 \\
\hline
\end{tabular}
\caption{Error estimation in WKB method for the noncommutative case $am=-0.1$ with  $n=0$, $M=1\ (R=2)$ and $\ell=2$.
	Orders for which the error exceeded the obtained value were omitted.
	} \label{tab5}
\end{table}

From the table we conclude that the optimal WKB order for axial case is 6, whereas for the polar case it is 4. The relative error between the WKB orders is of the order $10^{-4}$ in axial and $10^{-2}$ for the polar case.  \\ \\ \\ \\ \\
%%%%%%%%%%%%%%%%%%%%%%%%%%%%%%%%%%%%%%%%%%%%%%
%%%%%%%%%%%%%%%%%%%%%%%%%%%%%%%%%%%%%%%%%%%%%%%%
For the noncommutative case $am=-0.01$ with $\ell=2$ we obtain
\begin{table}[h!]
\begin{tabular}{m{7em} m{12em} m {8em} m{12em} m{8em}}
\hline
\\
WKB  order & Axial  QNM  & Error &  Polar  QNM   &  Error\\
 \\
\hline
\\
 12 & 0.373989 - 0.090051 i & 0.001792 & 0.372802 - 0.090057 i & 0.001604 \\
 11 & 0.374621 - 0.089899 i & 0.00062  & 0.374355 - 0.089683 i & 0.000877 \\
 10 & 0.374375 - 0.088872 i & 0.00055  & 0.374187 - 0.088979 i & 0.000427 \\
 9  & 0.374078 - 0.088942 i & 0.000181 & 0.373747 - 0.089084 i & 0.000255 \\
 8  & 0.374033 - 0.088753 i & 0.000159 & 0.373802 - 0.089315 i & 0.000143 \\
 7  & 0.373788 - 0.088811 i & 0.000133 & 0.373958 - 0.089278 i & 0.000092 \\
 6  & 0.373807 - 0.088892 i & {\bf 0.000072} & 0.373937 - 0.089190 i & 0.000059 \\
 5  & 0.373691 - 0.088919 i & 0.000119 & 0.373864 - 0.089207 i & {\bf 0.000038} \\
 4  & 0.373739 - 0.089121 i & 0.000228 & 0.373867 - 0.089218 i & 0.000324 \\
 3  & 0.373345 - 0.089215 i & 0.002501 & 0.373236 - 0.089369 i & 0.002636 \\
 \\
\hline
\end{tabular}
\caption{Error estimation in WKB method for the noncommutative case $am=-0.01$ with  $n=0$, $M=1\ (R=2)$ and $\ell=2$.} \label{tab6}
\end{table}

From the table we conclude that the optimal WKB order for the axial case is 6, whereas for the polar case it is 5. The relative error between the WKB orders is of the order $10^{-4}$ in both cases. 
%%%%%%%%%%%%%%%%%%%%%%%%%%%%%%%%%%%%%%%%%%%%%%
%%%%%%%%%%%%%%%%%%%%%%%%%%%%%%%%%%%%%%%%%%%%%%%%
\newpage
For the noncommutative case $am=-0.001$ with $\ell=2$ we obtain
\begin{table}[h!]
\begin{tabular}{m{7em} m{12em} m {8em} m{12em} m{8em}}
\hline
\\
WKB  order & Axial  QNM  & Error &  Polar  QNM   &  Error\\
 \\
\hline
\\
 12 & 0.373920 - 0.090036 i & 0.001903 & 0.372679 - 0.089851 i & 0.001724 \\
 11 & 0.374482 - 0.089901 i & 0.000611 & 0.374150 - 0.089498 i & 0.000853 \\
 10 & 0.374233 - 0.088855 i & 0.000562 & 0.373967 - 0.088732 i & 0.00045 \\
 9  & 0.373912 - 0.088931 i & 0.00019  & 0.373541 - 0.088833 i & 0.000254 \\
 8  & 0.373868 - 0.088747 i & 0.00016  & 0.373601 - 0.089083 i & 0.00015 \\
 7  & 0.373618 - 0.088806 i & 0.000136 & 0.373753 - 0.089046 i & 0.000092 \\
 6  & 0.373638 - 0.088891 i & {\bf 0.000073} & 0.373730 - 0.088950 i & 0.000062 \\
 5  & 0.373523 - 0.088919 i & 0.000121 & 0.373658 - 0.088967 i & {\bf 0.000038} \\
 4  & 0.373572 - 0.089124 i & 0.000227 & 0.373663 - 0.088986 i & 0.000323 \\
 3  & 0.373180 - 0.089217 i & 0.002501 & 0.373034 - 0.089135 i & 0.002623 \\
 \\
\hline
\end{tabular}
\caption{Error estimation in WKB method for the noncommutative case $am=-0.001$ with  $n=0$, $M=1\ (R=2)$ and $\ell=2$.} \label{tab7}
\end{table}

From the table we conclude that the optimal WKB order for the axial case is 6, whereas for the polar case it is 5. The relative error between the WKB orders is of the order $10^{-4}$ in both cases.  \\ \\ \\ \\ \\
%%%%%%%%%%%%%%%%%%%%%%%%%%%%%%%%%%%%%%%%%%%%%%
%%%%%%%%%%%%%%%%%%%%%%%%%%%%%%%%%%%%%%%%%%%%%%%%
For the noncommutative case $am=0.001$ with $\ell=2$ we obtain
\begin{table}[h!]
\begin{tabular}{m{7em} m{12em} m {8em} m{12em} m{8em}}
\hline
\\
WKB  order & Axial  QNM  & Error &  Polar  QNM   &  Error\\
 \\
\hline
\\
 12 & 0.373906 - 0.090033 i & 0.001929 & 0.372631 - 0.089801 i & 0.00173 \\
 11 & 0.374452 - 0.089901 i & 0.000609 & 0.374107 - 0.089447 i & 0.000856 \\
 10 & 0.374201 - 0.088851 i & 0.000565 & 0.373924 - 0.088678 i & 0.000452 \\
 9  & 0.373876 - 0.088929 i & 0.000192 & 0.373496 - 0.088780 i & 0.000255 \\
 8  & 0.373832 - 0.088745 i & 0.00016  & 0.373556 - 0.089030 i & 0.000151 \\
 7  & 0.373580 - 0.088805 i & 0.000137 & 0.373708 - 0.088994 i & 0.000093 \\
 6  & 0.373601 - 0.088891 i & {\bf 0.000074} & 0.373685 - 0.088897 i & 0.000062 \\
 5  & 0.373486 - 0.088918 i & 0.000121 & 0.373613 - 0.088914 i & {\bf 0.000038} \\
 4  & 0.373535 - 0.089124 i & 0.000227 & 0.373618 - 0.088933 i & 0.000322 \\
 3  & 0.373144 - 0.089218 i & 0.002502 & 0.372991 - 0.089083 i & 0.002621 \\
 \\
\hline
\end{tabular}
\caption{Error estimation in WKB method for the noncommutative case $am=0.001$ with $n=0$, $M=1\ (R=2)$ and $\ell=2$.} \label{tab8}
\end{table}

From the table we conclude that the optimal WKB order for the axial case is 6, whereas for the polar case it is 5. The relative error between the WKB orders is of the order $10^{-4}$ in both cases. 
%%%%%%%%%%%%%%%%%%%%%%%%%%%%%%%%%%%%%%%%%%%%%%
%%%%%%%%%%%%%%%%%%%%%%%%%%%%%%%%%%%%%%%%%%%%%%%%
\newpage
For the noncommutative case $am=0.01$ with $\ell=2$ we obtain
\begin{table}[h!]
\begin{tabular}{m{7em} m{12em} m {8em} m{12em} m{8em}}
\hline
\\
WKB  order & Axial  QNM  & Error &  Polar  QNM   &  Error\\
 \\
\hline
\\
 12 & 0.373840 - 0.090011 i & 0.002053 & 0.372415 - 0.089579 i & 0.001753 \\
 11 & 0.374316 - 0.089897 i & 0.0006   & 0.373918 - 0.089218 i & 0.000872 \\
 10 & 0.374061 - 0.088832 i & 0.000577 & 0.373731 - 0.088434 i & 0.000461 \\ 
 9  & 0.373710 - 0.088915 i & 0.000203 & 0.373296 - 0.088537 i & 0.00026 \\
 8  & 0.373668 - 0.088735 i & 0.000161 & 0.373357 - 0.088795 i & 0.000155 \\
 7  & 0.373412 - 0.088796 i & 0.00014  & 0.373513 - 0.088758 i & 0.000095 \\
 6  & 0.373433 - 0.088888 i & {\bf 0.000076} & 0.373489 - 0.088657 i & 0.000065 \\
 5  & 0.373318 - 0.088915 i & 0.000122 & 0.373413 - 0.088675 i & {\bf 0.000041} \\
 4  & 0.373368 - 0.089124 i & 0.000226 & 0.373418 - 0.088696 i & 0.000318 \\
 3  & 0.372981 - 0.089216 i & 0.002503 & 0.372801 - 0.088843 i & 0.002613 \\
 \\
\hline
\end{tabular}
\caption{Error estimation in WKB method for the noncommutative case $am=0.01$ with  $n=0$, $M=1 (R=2)$ and $\ell=2$} \label{tab9}
\end{table}

From the table we conclude that the optimal WKB order for the axial case is 6, whereas for the polar case, it is 5. The relative error between the WKB orders is of the order $10^{-4}$ in both cases.  \\ \\ \\ \\ \\
%%%%%%%%%%%%%%%%%%%%%%%%%%%%%%%%%%%%%%%%%%%%%%
%%%%%%%%%%%%%%%%%%%%%%%%%%%%%%%%%%%%%%%%%%%%%%%%
For the noncommutative case $am=0.1$ with $\ell=2$ we obtain
\begin{table}[h!]
\begin{tabular}{m{7em} m{12em} m {8em} m{12em} m{8em}}
\hline
\\
WKB  order & Axial  QNM  & Error &  Polar  QNM   &  Error\\
 \\
\hline
\\
 12 &  &                               & 0.369608 - 0.087506 i & 0.002105 \\
 11 &  &                               & 0.372146 - 0.086909 i & 0.0014 \\
 10 & 0.423526 - 0.090112 i & 0.116517 & 0.371917 - 0.085922 i & 0.000538 \\  
 9  & 0.371627 - 0.102697 i & 0.027848 & 0.371565 - 0.086003 i & 0.000249 \\  
 8  & 0.367866 - 0.088119 i & 0.007773 & 0.371643 - 0.086338  i & 0.000221\\  
 7  & 0.371941 - 0.087153 i & 0.002263 & 0.371912 - 0.086276 i & 0.000148 \\
 6  & 0.372337 - 0.088826 i & 0.00094  & 0.371888 - 0.086175 i & {\bf 0.00006} \\
 5  & 0.371603 - 0.089001 i & 0.000379 & 0.371832 - 0.086188 i & 0.000063 \\
 4  & 0.371587 - 0.088938 i & {\bf 0.000085} & 0.371857 - 0.086298 i & 0.000283 \\
 3  & 0.371435 - 0.088975 i & 0.002594 & 0.371317 - 0.086423 i & 0.002616 \\
 \\
\hline
\end{tabular}
\caption{Error estimation in WKB method for the noncommutative case $am=0.1$ with $n=0$, $M=1\ (R=2)$ and $\ell=2$.
	Orders for which the error exceeded the obtained value were omitted.
	} \label{tab10}
\end{table}

From the table we conclude that the optimal WKB order for the axial case is 4, whereas for the polar case it is 6. The relative error between the WKB orders is of the order $10^{-4}$ in both cases. We observe that the higher-order WKB method breaks down after reaching certain orders for axial case. This phenomenon appears to be a generic feature, particularly noticeable for higher (positive) values of the noncommutative parameter $am$ for the axial case.
%%%%%%%%%%%%%%%%%%%%%%%%%%%%%%%%%%%%%%%%%%%%%%
%%%%%%%%%%%%%%%%%%%%%%%%%%%%%%%%%%%%%%%%%%%%%%%%
\newpage
For the noncommutative case $am=0.2$ with $\ell=2$ we obtain
\begin{table}[h!]
\begin{tabular}{m{7em} m{12em} m {8em} m{12em} m{8em}}
\hline
\\
WKB  order & Axial  QNM  & Error &  Polar  QNM   &  Error\\
 \\
\hline
\\
 12 &   &                               & 0.371439 - 0.084352 i & 0.001621 \\
 11 &   &                               & 0.372232 - 0.084172 i & 0.000735 \\
 10 &   &                               & 0.371963 - 0.082977 i & 0.00081 \\
 9  &   &                               & 0.370930 - 0.083209 i & 0.000582 \\
 8  &   &                               & 0.371035 - 0.083679 i & 0.000246 \\
 7  & 0.375044 - 0.026050 i & 0.207929  & 0.371129 - 0.083658 i & {\bf 0.000056} \\   
 6  & 0.384557 - 0.088907 i & 0.03337   & 0.371116 - 0.083600 i & 0.000109 \\  
 5  & 0.369412 - 0.092552 i & 0.008114  & 0.370910 - 0.083647 i & 0.000152 \\
 4  & 0.368345 - 0.088195 i & {\bf 0.002378}  & 0.370959 - 0.083862 i & 0.000356 \\
 3  & 0.369909 - 0.087822 i & 0.003151  & 0.370299 - 0.084011 i & 0.002625 \\
 \\
\hline
\end{tabular}
\caption{Error estimation in WKB method for the noncommutative case $am=0.2$ with  $n=0$, $M=1\ (R=2)$ and $\ell=2$.	
	Orders for which the error exceeded the obtained value were omitted.
	} \label{tab11}
\end{table}

From the table we conclude that the optimal WKB order for the axial case is 4, whereas for the polar case it is 7. The relative error between the WKB orders is of the order $10^{-2}$ in the axial case and $10^{-4}$ in the polar case.  \\ \\ \\ \\ \\
%%%%%%%%%%%%%%%%%%%%%%%%%%%%%%%%%%%%%%%%%%%%%%
%%%%%%%%%%%%%%%%%%%%%%%%%%%%%%%%%%%%%%%%%%%%%%%%
Now we consider the mode with $\ell=3$. For the commutative case $(am=0)$ we obtain
%%%%%%%%%%%%%%%%%%%%%%%%%%%%%%%%%%%%%%%%%%%%%%%%
\begin{table}[h!]
\begin{tabular}{m{7em} m{12em} m {8em} m{12em} m{8em}}
\hline
\\
WKB  order & Axial  QNM  & Error &  Polar  QNM   &  Error\\
 \\
\hline
\\
 12 & 0.599443 - 0.092703 i & 2.21 $\times 10^{-7}$ & 0.599443 - 0.092703 i & 1.898 $\times 10^{-7}$ \\
 11 & 0.599443 - 0.092703 i & 3.92 $\times 10^{-7}$ & 0.599443 - 0.092703 i & 9.85 $\times 10^{-8}$ \\
 10 & 0.599443 - 0.092703 i & 4.45 $\times 10^{-7}$ & 0.599443 - 0.092703 i & 4.03 $\times 10^{-8}$ \\
 9  & 0.599444 - 0.092703 i & 3.59 $\times 10^{-7}$ & 0.599443 - 0.092703 i & 7.84 $\times 10^{-8}$ \\
 8  & 0.599443 - 0.092703 i & 1.94 $\times 10^{-7}$ & 0.599443 - 0.092703 i & 1.271 $\times 10^{-7}$ \\
 7  & 0.599443 - 0.092703 i & 2.38 $\times 10^{-7}$ & 0.599443 - 0.092703 i & 2.18 $\times 10^{-7}$ \\
 6  & 0.599443 - 0.092703 i & 1.081$\times 10^{-6}$ & 0.599443 - 0.092703 i & 8.361 $\times 10^{-7}$ \\
 5  & 0.599441 - 0.092703 i & 1.347$\times 10^{-6}$ & 0.599442 - 0.092703 i & 1.5108 $\times 10^{-6}$ \\
 4  & 0.599441 - 0.092701 i & 0.000089              & 0.599441 - 0.092700 i & 0.000089 \\
 3  & 0.599265 - 0.092728 i & 0.001116              & 0.599264 - 0.092728 i & 0.001117 \\
 \\
\hline
\end{tabular}
\caption{Error estimation in WKB method for the commutative case $(am=0)$ with  $n=0$, $M=1\ (R=2)$ and $\ell=3$.} \label{tab12}
\end{table}

From the table we conclude that the optimal WKB order for the axial case is 12, whereas for the polar case it is 10. The relative error between the WKB orders is negligible, $10^{-7}$ for the axial case and $10^{-8}$ for the polar case. 
%%%%%%%%%%%%%%%%%%%%%%%%%%%%%%%%%%%%%%%%%%%%%%
%%%%%%%%%%%%%%%%%%%%%%%%%%%%%%%%%%%%%%%%%%%%%%%%
\newpage
For the noncommutative case $am=-0.2$ with $\ell=3$ we obtain
\begin{table}[h!]
\begin{tabular}{m{7em} m{12em} m {8em} m{12em} m{8em}}
\hline
\\
WKB  order & Axial  QNM  & Error &  Polar  QNM   &  Error\\
 \\
\hline
\\
 12 & 0.602760 - 0.092375 i & 4.23 $\times 10^{-6}$ &  \\
 11 & 0.602764 - 0.092374 i & 2.93 $\times 10^{-6}$ &  \\
 10 & 0.602765 - 0.092379 i & 2.66 $\times 10^{-6}$ &  \\
 9  & 0.602768 - 0.092378 i & 2.11 $\times 10^{-6}$ &  \\
 8  & 0.602768 - 0.092381 i &1.86 $\times 10^{-6}$  &  \\
 7  & 0.602770 - 0.092381 i & 2.61 $\times 10^{-6}$ &  \\
 6  & 0.602771 - 0.092385 i & 2.45 $\times 10^{-6}$ & 0.649712 - 0.099181 i & 0.114998 \\
 5  & 0.602772 - 0.092385 i & 7.1  $\times 10^{-6}$ & 0.601994 - 0.107043 i & 0.025139 \\
 4  & 0.602774 - 0.092399 i & 0.000089              & 0.599747 - 0.093586 i & 0.007185 \\
 3  & 0.602598 - 0.092426 i & 0.001166              & 0.604367 - 0.092871 i & 0.003341 \\
 \\
\hline
\end{tabular}
\caption{Error estimation in WKB method for the noncommutative case $am=-0.2$ with  $n=0$, $M=1\ (R=2)$ and $\ell=3$.} \label{tab13}
\end{table}

From the table we conclude that the optimal WKB order is 8 for both the axial case and 3 for the polar case. The relative error between the WKB orders is of the order $10^{-6}$ in axial case and $10^{-2}$ in polar case.  \\ \\ \\ \\ \\
%%%%%%%%%%%%%%%%%%%%%%%%%%%%%%%%%%%%%%%%%%%%%%
%%%%%%%%%%%%%%%%%%%%%%%%%%%%%%%%%%%%%%%%%%%%%%%%
For the noncommutative case $am=-0.1$ with $\ell=3$ we obtain
\begin{table}[h!]
\begin{tabular}{m{7em} m{12em} m {8em} m{12em} m{8em}}
\hline
\\
WKB  order & Axial  QNM  & Error &  Polar  QNM   &  Error\\
 \\
\hline
\\
 12 & 0.600921 - 0.092632 i & 1.226 $\times 10^{-6}$ &  \\
 11 & 0.600920 - 0.092633 i & 9.68 $\times 10^{-7}$  &  0.637566 + 0.066313 i & 0.502846 \\
 10 & 0.600920 - 0.092632 i & 5.79 $\times 10^{-7}$  &  0.641747 - 0.098722 i & 0.088089 \\
 9  & 0.600920 - 0.092631 i & 8.59 $\times 10^{-7}$  &  0.599441 - 0.105690 i & 0.022373 \\
 8  & 0.600920 - 0.092630 i & 9.28 $\times 10^{-7}$  &  0.597353 - 0.093123 i & 0.006679 \\
 7  & 0.600921 - 0.092630 i & 8.03 $\times 10^{-7}$  &  0.601463 - 0.092486 i & 0.002223 \\
 6  & 0.600920 - 0.092628 i & 9.56 $\times 10^{-7}$  &  0.601704 - 0.094037 i & 0.000862 \\
 5  & 0.600922 - 0.092628 i & 6.71 $\times 10^{-7}$  &  0.601002 - 0.094147 i & 0.000406 \\
 4  & 0.600922 - 0.092628 i & 0.000084               &  0.600941 - 0.093756 i & 0.000203 \\
 3  & 0.600755 - 0.092653 i & 0.001134               &  0.601035 - 0.093742 i & 0.001263 \\
 \\
\hline
\end{tabular}
\caption{Error estimation in WKB method for the noncommutative case $am=-0.1$ with  $n=0$, $M=1\ (R=2)$ and $\ell=3$.} \label{tab14}
\end{table}

From the table we conclude that the optimal WKB order for the axial case is 10, whereas for the polar case it is 4. The relative error between the WKB orders is of the order $10^{-7}$ for the axial and $10^{-3}$ for the polar case. 
%%%%%%%%%%%%%%%%%%%%%%%%%%%%%%%%%%%%%%%%%%%%%%
%%%%%%%%%%%%%%%%%%%%%%%%%%%%%%%%%%%%%%%%%%%%%%%%
\newpage
For the noncommutative case $am=-0.01$ with $\ell=3$ we obtain
\begin{table}[h!]
\begin{tabular}{m{7em} m{12em} m {8em} m{12em} m{8em}}
\hline
\\
WKB  order & Axial  QNM  & Error &  Polar  QNM   &  Error\\
 \\
\hline
\\
 12 & 0.599573 - 0.092706 i & 2.17 $\times 10^{-7}$  & 0.599571 - 0.092828 i & 4.73 $\times 10^{-7}$ \\
 11 & 0.599572 - 0.092706 i & 3.77 $\times 10^{-7}$  & 0.599571 - 0.092827 i & 2.02 $\times 10^{-7}$ \\
 10 & 0.599572 - 0.092707 i & 4.28 $\times 10^{-7}$  & 0.599571 - 0.092828 i & 2.22 $\times 10^{-7}$ \\
 9  & 0.599573 - 0.092707 i & 3.47 $\times 10^{-7}$  & 0.599571 - 0.092828 i & 1.47 $\times 10^{-7}$ \\
 8  & 0.599573 - 0.092706 i & 1.84 $\times 10^{-7}$  & 0.599571 - 0.092828 i & 2.09 $\times 10^{-7}$ \\
 7  & 0.599573 - 0.092706 i & 2.12 $\times 10^{-7}$  & 0.599571 - 0.092828 i & 2.2 $\times 10^{-7}$ \\
 6  & 0.599573 - 0.092706 i & 1.078 $\times 10^{-6}$ & 0.599571 - 0.092828 i & 9.68 $\times 10^{-7}$ \\
 5  & 0.599571 - 0.092706 i & 1.416 $\times 10^{-6}$ & 0.599569 - 0.092828 i & 1.812 $\times 10^{-6}$ \\
 4  & 0.599570 - 0.092704 i & 0.000088               & 0.599569 - 0.092825 i & 0.000089 \\
 3  & 0.599395 - 0.092732 i & 0.001116               & 0.599392 - 0.092852 i & 0.001119 \\
 \\
\hline
\end{tabular}
\caption{Error estimation in WKB method for the noncommutative case $am=-0.01$ with  $n=0$, $M=1\ (R=2)$ and $\ell=3$.} \label{tab15}
\end{table}

From the table we conclude that the optimal WKB order for the axial case is 8, whereas for the polar case it is 9. The relative error between the WKB orders is of the order $10^{-7}$ for both cases.  \\ \\ \\ \\ \\
%%%%%%%%%%%%%%%%%%%%%%%%%%%%%%%%%%%%%%%%%%%%%%
%%%%%%%%%%%%%%%%%%%%%%%%%%%%%%%%%%%%%%%%%%%%%%%%
For the noncommutative case $am=-0.001$ with $\ell=3$ we obtain
\begin{table}[h!]
\begin{tabular}{m{7em} m{12em} m {8em} m{12em} m{8em}}
\hline
\\
WKB  order & Axial  QNM  & Error &  Polar  QNM   &  Error\\
 \\
\hline
\\
 12 & 0.599456 - 0.092703 i & 2.2 $\times 10^{-7}$   & 0.599456 - 0.092715 i & 1.889 $\times 10^{-7}$ \\
 11 & 0.599456 - 0.092703 i & 3.91 $\times 10^{-7}$  & 0.599456 - 0.092716 i & 9.77 $\times 10^{-8}$ \\
 10 & 0.599456 - 0.092704 i & 4.43 $\times 10^{-7}$  & 0.599456 - 0.092716 i & 4.04 $\times 10^{-8}$ \\
 9  & 0.599456 - 0.092704 i & 3.58 $\times 10^{-7}$  & 0.599456 - 0.092716 i & 7.86 $\times 10^{-8}$  \\
 8  & 0.599456 - 0.092703 i & 1.93 $\times 10^{-7}$  & 0.599456 - 0.092716 i & 1.275 $\times 10^{-7}$ \\
 7  & 0.599456 - 0.092703 i & 2.36 $\times 10^{-7}$  & 0.599456 - 0.092716 i & 2.175 $\times 10^{-7}$ \\
 6  & 0.599456 - 0.092703 i & 1.078 $\times 10^{-6}$ & 0.599456 - 0.092715 i & 8.363 $\times 10^{-7}$ \\
 5  & 0.599454 - 0.092703 i & 1.347 $\times 10^{-6}$ & 0.599454 - 0.092716 i & 1.515 $\times 10^{-6}$ \\
 4  & 0.599454 - 0.092702 i & 0.000089               & 0.599454 - 0.092713 i & 0.000089 \\
 3  & 0.599278 - 0.092729 i & 0.001116               & 0.599277 - 0.092740 i & 0.001117 \\
 \\
\hline
\end{tabular}
\caption{Error estimation in WKB method for the noncommutative case $am=-0.001$ with  $n=0$, $M=1\ (R=2)$ and $\ell=3$.} \label{tab16}
\end{table}

From the table we conclude that the optimal WKB order for the axial case is 8, whereas for the polar case it is 10. The relative error between the WKB orders is of the order $10^{-7}$ for axial and $10^{-8}$ for the polar case. 
%%%%%%%%%%%%%%%%%%%%%%%%%%%%%%%%%%%%%%%%%%%%%%
%%%%%%%%%%%%%%%%%%%%%%%%%%%%%%%%%%%%%%%%%%%%%%%%
\newpage
For the noncommutative case $am=0.001$ with $\ell=3$ we obtain
\begin{table}[h!]
\begin{tabular}{m{7em} m{12em} m {8em} m{12em} m{8em}}
\hline
\\
WKB  order & Axial  QNM  & Error &  Polar  QNM   &  Error\\
 \\
\hline
\\
 12 & 0.599431 - 0.092702 i & 2.22 $\times 10^{-7}$  & 0.599431 - 0.092690 i & 1.908 $\times 10^{-7}$ \\
 11 & 0.599430 - 0.092702 i & 3.93 $\times 10^{-7}$  & 0.599431 - 0.092690 i & 9.93 $\times 10^{-8}$ \\
 10 & 0.599430 - 0.092703 i & 4.47 $\times 10^{-7}$  & 0.599431 - 0.092690 i & 4.05 $\times 10^{-8}$ \\
 9  & 0.599431 - 0.092703 i & 3.61 $\times 10^{-7}$  & 0.599431 - 0.092690 i & 7.8 $\times 10^{-8}$ \\
 8  & 0.599431 - 0.092702 i & 1.95 $\times 10^{-7}$  & 0.599431 - 0.092691 i & 1.271 $\times 10^{-7}$ \\
 7  & 0.599431 - 0.092702 i & 2.39 $\times 10^{-7}$  & 0.599431 - 0.092691 i & 2.178 $\times 10^{-7}$ \\
 6  & 0.599431 - 0.092702 i & 1.084 $\times 10^{-6}$ & 0.599431 - 0.092690 i & 8.375 $\times 10^{-7}$ \\
 5  & 0.599429 - 0.092702 i & 1.348 $\times 10^{-6}$ & 0.599429 - 0.092690 i & 1.5105 $\times 10^{-6}$ \\
 4  & 0.599428 - 0.092701 i & 0.000089               & 0.599429 - 0.092688 i & 0.000089 \\
 3  & 0.599252 - 0.092728 i & 0.001116               & 0.599252 - 0.092715 i & 0.001117 \\
 \\
\hline
\end{tabular}
\caption{Error estimation in WKB method for the noncommutative case $am=0.001$ with  $n=0$, $M=1\ (R=2)$ and $\ell=3$.} \label{tab17}
\end{table}

From the table we conclude that the optimal WKB order for the axial case is 8, whereas for the polar case it is 10. The relative error between the WKB orders is of the order $10^{-7}$ for the axial and $10^{-8}$ for the polar case.  \\ \\ \\ \\ \\
%%%%%%%%%%%%%%%%%%%%%%%%%%%%%%%%%%%%%%%%%%%%%%
%%%%%%%%%%%%%%%%%%%%%%%%%%%%%%%%%%%%%%%%%%%%%%%%
For the noncommutative case $am=0.01$ with $\ell=3$ we obtain
\begin{table}[h!]
\begin{tabular}{m{7em} m{12em} m {8em} m{12em} m{8em}}
\hline
\\
WKB  order & Axial  QNM  & Error &  Polar  QNM   &  Error\\
 \\
\hline
\\
 12 & 0.599318 - 0.092696 i & 1.73 $\times 10^{-7}$  & 0.599324 - 0.092577 i & 2.061 $\times 10^{-7}$ \\
 11 & 0.599318 - 0.092697 i & 4.17 $\times 10^{-7}$  & 0.599324 - 0.092577 i & 1.023 $\times 10^{-7}$ \\
 10 & 0.599318 - 0.092697 i & 4.73 $\times 10^{-7}$  & 0.599324 - 0.092577 i & 3.3 $\times 10^{-8}$ \\
 9  & 0.599319 - 0.092697 i & 3.66 $\times 10^{-7}$  & 0.599324 - 0.092577 i & 8.51 $\times 10^{-8}$ \\
 8  & 0.599319 - 0.092697 i & 2.37 $\times 10^{-7}$  & 0.599324 - 0.092577 i & 1.28 $\times 10^{-7}$ \\
 7  & 0.599319 - 0.092697 i & 2.03 $\times 10^{-7}$  & 0.599324 - 0.092577 i & 2.08 $\times 10^{-7}$ \\
 6  & 0.599318 - 0.092696 i & 1.177 $\times 10^{-6}$ & 0.599324 - 0.092577 i & 8.893 $\times 10^{-7}$ \\
 5  & 0.599316 - 0.092697 i & 1.476 $\times 10^{-6}$ & 0.599322 - 0.092577 i & 1.6208 $\times 10^{-6}$ \\
 4  & 0.599316 - 0.092695 i & 0.000088               & 0.599322 - 0.092574 i & 0.000089 \\
 3  & 0.599140 - 0.092722 i & 0.001117               & 0.599146 - 0.092601 i & 0.001116 \\
 \\
\hline
\end{tabular}
\caption{Error estimation in WKB method for the noncommutative case $am=0.01$ with $n=0$, $M=1\ (R=2)$ and $\ell=3$.} \label{tab18}
\end{table}

From the table we conclude that the optimal WKB order for the axial case is 12, whereas for the polar case it is 10. The relative error between the WKB orders is of the order $10^{-7}$ in the axial case and $10^{-8}$ in the polar case. 
%%%%%%%%%%%%%%%%%%%%%%%%%%%%%%%%%%%%%%%%%%%%%%
%%%%%%%%%%%%%%%%%%%%%%%%%%%%%%%%%%%%%%%%%%%%%%%%
\newpage
For the noncommutative case $am=0.1$ with $\ell=3$ we obtain
\begin{table}[h!]
\begin{tabular}{m{7em} m{12em} m {8em} m{12em} m{8em}}
\hline
\\
WKB  order & Axial  QNM  & Error &  Polar  QNM   &  Error\\
 \\
\hline
\\
 12 & 0.552799 - 0.086858 i & 0.087919 & 0.598585 - 0.091404 i & 3.0149 $\times 10^{-6}$ \\
 11 & 0.600442 - 0.079966 i & 0.024979 & 0.598587 - 0.091404 i & 1.4565 $\times 10^{-6}$ \\
 10 & 0.602350 - 0.093218 i & 0.007056 & 0.598587 - 0.091402 i & 1.466  $\times 10^{-6}$ \\
 9  & 0.598097 - 0.093881 i & 0.002289 & 0.598589 - 0.091402 i & 1.132 $\times 10^{-6}$ \\
 8  & 0.597857 - 0.092341 i & 0.000842 & 0.598589 - 0.091401 i & 1.5243 $\times 10^{-6}$ \\
 7  & 0.598489 - 0.092243 i & 0.000354 & 0.598592 - 0.091400 i & 1.485 $\times 10^{-6}$ \\
 6  & 0.598535 - 0.092542 i & 0.000175 & 0.598592 - 0.091400 i & 1.4303 $\times 10^{-6}$ \\
 5  & 0.598362 - 0.092569 i & 0.000105 & 0.598595 - 0.091400 i & 2.6585 $\times 10^{-6}$ \\
 4  & 0.598344 - 0.092452 i & 0.00007  & 0.598596 - 0.091404 i & 0.000083  \\
 3  & 0.598269 - 0.092464 i & 0.001177 & 0.598431 - 0.091429 i & 0.001136  \\
 \\
\hline
\end{tabular}
\caption{Error estimation in WKB method for the noncommutative case $am=0.1$ with  $n=0$, $M=1\ (R=2)$ and $\ell=3$} \label{tab19}
\end{table}

From the table we conclude that the optimal WKB order is 4 for axial case and 9 for polar case. The relative error between the WKB orders is of the order $10^{-4}$ in the axial and $10^{-6}$ in the polar case.  \\ \\ \\ \\ \\
%%%%%%%%%%%%%%%%%%%%%%%%%%%%%%%%%%%%%%%%%%%%%%
%%%%%%%%%%%%%%%%%%%%%%%%%%%%%%%%%%%%%%%%%%%%%%%%
For the noncommutative case $am=0.2$ with $\ell=3$ we obtain
\begin{table}[h!]
\begin{tabular}{m{7em} m{12em} m {8em} m{12em} m{8em}}
\hline
\\
WKB  order & Axial  QNM  & Error &  Polar  QNM   &  Error\\
 \\
\hline
\\
 12 &  &                                & 0.598413 - 0.090093 i & 5.2962 $\times 10^{-6}$ \\
 11 &  &                                & 0.598405 - 0.090094 i & 4.9559 $\times 10^{-6}$ \\
 10 &  &                                & 0.598405 - 0.090099 i & 3.3432 $\times 10^{-6}$ \\
 9  &  &                                & 0.598401 - 0.090100 i & 3.3601 $\times 10^{-6}$ \\
 8  &  &                                & 0.598402 - 0.090106 i & 2.9334 $\times 10^{-6}$ \\
 7  & 0.600775 - 0.073159 i & 0.042215 & 0.598400 - 0.090106 i & 3.6448 $\times 10^{-6}$ \\
 6  & 0.603498 - 0.092906 i & 0.010492 & 0.598401 - 0.090113 i & 4.6264 $\times 10^{-6}$ \\
 5  & 0.597349 - 0.093862 i & 0.003336 & 0.598395 - 0.090114 i & 0.000011 \\
 4  & 0.596979 - 0.091480 i & 0.00131  & 0.598399 - 0.090136 i & 0.000098 \\
 3  & 0.597999 - 0.091324 i & 0.001636 & 0.598206 - 0.090165 i & 0.001163 \\
 \\
\hline
\end{tabular}
\caption{Error estimation in WKB method for the noncommutative case $am=0.2$ with  $n=0$, $M=1\ (R=2)$ and $\ell=3$.
	Orders for which the error exceeded the obtained value were omitted.
	} \label{tab20}
\end{table}

From the table we conclude that the optimal WKB order for the axial case is 4, whereas for the polar case it is 8. The relative error between the WKB orders is of the order $10^{-3}$ in the axial case and $10^{-6}$ in the polar case. As observed earlier, for the axial modes, the higher-order WKB breaks for large positive $am$ values.
%%%%%%%%%%%%%%%%%%%%%%%%%%%%%%%%%%%%%%%%%%%%%%
%%%%%%%%%%%%%%%%%%%%%%%%%%%%%%%%%%%%%%%%%%%%%%%%

\pagebreak

The QNM calculation using the higher-order WKB method can be summarised as follows:
\begin{enumerate}
\item  For $\ell=2$, the optimal WKB order for the commutative case is 6/5 for the axial/polar perturbations. For negative (positive) $am$ values, the optimal order for the axial (polar) is around the order in the commutative case. However, for large positive (negative) $am$ values ($|am| \sim 0.1$ or $|am|>0.1$), the optimal orders are lower for the axial (polar) case. 
\item  For all $\ell$ values, in the acceptable WKB orders, the noncommutative corrections are larger than the errors.
\item  For $\ell>2$, the errors in the higher-order WKB are negligible. However, the issue with positive (negative) $am$ for axial (polar) mode persists.
\end{enumerate}

\bibliography{BibTeX}

\end{document}